\documentclass[useAMS,usenatbib,usegraphicx]{mn2e}
\bibpunct[,]{(}{)}{;}{a}{}{}

\newcommand{\hm}{h^{-1}}

\usepackage{amssymb}

\title[ICL formation in hierarchical galaxy formation models]
{On the formation and physical properties of the Intra-Cluster Light in 
hierarchical galaxy formation models}
\author[E.~Contini et al.]
        {E.~Contini,$^{1,2}$\thanks{Email: contini@oats.inaf.it} 
        G.~De Lucia,$^{2}$
	\'A.~Villalobos,$^{2}$
        S.~Borgani,$^{1,2,3}$
        \\      
        $^1$ Dipartimento di Astronomia, Universit{\' a} di Trieste, 
        via G.B. Tiepolo 11, I-34131 Trieste,Italy \\
	$^2$ INAF - Astronomical Observatory of Trieste, 
        via G.B. Tiepolo 11, I-34143 Trieste, Italy \\
	$^3$INFN, Sezione di Trieste, Via Valerio 2, I-34127 Trieste, Italy
        \\}

\pagerange{\pageref{firstpage}--\pageref{lastpage}}
\pubyear{2012}

\begin{document}

\maketitle

\label{firstpage}

\begin{abstract}
We study the formation of the Intra-Cluster Light (ICL) using a semi-analytic
model of galaxy formation, coupled to merger trees extracted from N-body simulations of
groups and clusters. We assume that the ICL forms by (1) stellar stripping of
satellite galaxies and (2) relaxation processes that take place during galaxy
mergers.  The fraction of ICL in groups and clusters predicted by our models
ranges between 10 and 40 per cent, with a large halo-to-halo scatter and no 
halo mass dependence. 
We note, however, that our predicted ICL fractions depend on the resolution: 
for a set of simulations with particle mass one order of magnitude larger than 
that adopted in the high resolution runs used in our study, we find that the predicted 
ICL fractions are ~30-40 per cent larger than those found in the high resolution runs.
On cluster scale, large part of the scatter is due to a range of dynamical
histories, while on smaller scale it is driven by individual
accretion events and stripping of very massive satellites, $M_{*} \gtrsim
10^{10.5} M_{\odot}$, that we find to be the major contributors to the ICL.
The ICL in our models forms very late (below $z\sim 1$), and a fraction varying
between 5 and 25 per cent of it has been accreted during the hierarchical
growth of haloes. In agreement with recent observational measurements, we find
the ICL to be made of stars covering a relatively large range of
metallicity, with the bulk of them being sub-solar. 
\end{abstract}


\section[]{Introduction}
\label{sec:intro}

The presence of a diffuse population of intergalactic stars in galaxy clusters
was first proposed by \citet{zwicky37}, and later confirmed by the same author
using observations of the Coma cluster with a 48-inch schmidt telescope
\citep{zwicky52}. More recent observational studies have confirmed that a
substantial fraction of stars in clusters are not bound to galaxies. This
diffuse component is generally referred to as Intra-Cluster Light (hereafter
ICL).

Both from the observational and the theoretical point of view, it is not
trivial to define the ICL component. A fraction of central cluster galaxies 
are characterized by a faint and extended
stellar halo. These galaxies are classified as \emph{cD galaxies}, where `c'
refers to the fact that these galaxies are very large and stands for 
supergiant and `D' for diffuse (\citealt{matthews64}), to highlight 
the presence of a diffuse stellar envelope made of stars that are not bound 
to the galaxy itself. Separating these two components is not an easy task. On the
observational side, some authors use an isophotal limit to cut off the 
light from satellite galaxies, while the distinction between the brightest 
cluster galaxy (hereafter BCG) and ICL is based on profile decomposition
\citep[e.g.][]{zibetti}. Others \citep[e.g.][]{gonzalez05} rely on
two-dimensional profile fittings to model the surface brightness profile of
brightest cluster galaxies. In the framework of numerical simulations,
additional information is available, and the ICL component has been defined
using either a binding energy definition \citep[i.e. all stars that are not
bound to identified galaxies, e.g.][]{giuseppe}, or variations of this
technique that take advantage of the dynamical information provided by the
simulations \citep[e.g.][]{Dolag_etal_2010}. In a recent work, \citet{rudick11}
discuss different methods that have been employed both for
observational and for theoretical data, and apply them to a suite of N-body
simulations of galaxy clusters. They find that different methods can change the
measured fraction\footnote{This is the ratio between the mass or luminosity in
  the ICL component and the total stellar mass or luminosity enclosed within
  some radius, usually $R_{200}$ or $R_{500}$. In this work, we will use
  $R_{200}$, defined as the radius that encloses a mean density of 200 times the
  critical density of the Universe at the redshift of interest.} of ICL by up
to a factor of about four (from $\sim 9$ to $\sim 36$ per cent). In contrast,
\cite{puchwein} apply four different methods to identify the ICL in
hydrodynamical SPH simulations of cluster galaxies, and consistently find a
significant ICL stellar fraction ($\sim$ 45 per cent).

There is no general agreement in the literature about how the ICL fraction
varies as a function of cluster mass. \citet{zibetti} find that richer clusters
(the richness being determined by the number of red-sequence galaxies), and
those with a more luminous BCG have brighter ICL than their
counterparts. However, they find roughly constant ICL fractions as a function 
of halo mass, within the
uncertainties and sample variance. In contrast, \citet{lin} empirically infer an
increasing fraction of ICL with increasing cluster mass. 
To estimate the amount of ICL, they use the observed correlation between 
the cluster luminosity and mass and a simple merger tree model for cluster formation.
Results are inconclusive also on the theoretical side, with claims of increasing ICL
fractions for more massive haloes
(e.g. \citealt{giuseppe04,purcell07,giuseppe,purcell08}), as well as findings
of no significant increase of the ICL fraction with cluster mass
(e.g. \citealt{pigi,henriques,puchwein}), at least for systems more massive
than $10^{13} \, M_{\odot}/h$.

Different physical mechanisms may be at play in the formation of the ICL, and
their relative importance can vary during the dynamical history of the
cluster. Stars can be stripped away from satellite galaxies orbiting within the
cluster, by tidal forces exerted either during interactions with other cluster
galaxies, or by the cluster potential. This is supported by observations of
arclets and similar tidal features that have been identified in the Coma, Centaurus 
and Hydra I clusters \citep{gregg, trentham, calcaneo, arnaboldi12}. As pointed out by several
authors, in a scenario where galaxy stripping and disruption are the main
mechanisms for the production of the ICL, the major contribution comes from
galaxies falling onto the cluster along almost radial orbits, since tidal
interactions by the cluster potential are strongest for these
galaxies. Numerical simulations have also shown that large amounts of ICL can
come from `pre-processing' in galaxy groups that are later accreted onto
massive clusters \citep{mihos,willman,rudick06,sommer}. In addition,
\citet{giuseppe} found that the formation of the ICL is tightly linked to the
build-up of the BCG and of the other
massive cluster galaxies, a scenario supported by other theoretical studies
(e.g. \citealt{diemand,abadi,font,read}).  It is important, however, to
consider that results from numerical simulations might be affected by numerical
problems. \citet{giuseppe} find an increasing fraction of ICL when
increasing the numerical resolution of their simulations. In addition,
\citet{puchwein} show that a significant fraction ($\sim 30$ per cent) of the
ICL identified in their simulations forms in gas clouds that were stripped from
the dark matter haloes of galaxies infalling onto the cluster. Fluid
instabilities, that are not well treated within the SPH framework, might be
able to disrupt these clouds suppressing this mode of ICL formation.

In this paper, we use the semi-analytic model presented in \citet[][ hereafter
  DLB07]{dlb}, that we extend by including three different prescriptions for
the formation of the ICL. We couple this model to a suite of high-resolution
N-body simulations of galaxy clusters to study the formation and evolution of
the ICL component, as well as its physical properties, and the influence of the
updated prescriptions on model basic predictions (in particular, the galaxy
stellar mass function, and the mass of the BCGs). 
There are some advantages in
using semi-analytic models to describe the ICL formation with respect to 
hydrodynamical simulations: they do not suffer from
numerical effects related to the fragility of poorly resolved galaxies, and
allow the relative influence of different channels of ICL generation to be clearly quantified.
However, the size and abundance of satellite galaxies (that influence the amount 
of predicted ICL) might be estimated incorrectly in these models. We will comment 
on these issues in the following.

The layout of the paper is as follows. In Section~\ref{sec:sim} we introduce
the simulations used in our study, and in Section~\ref{sec:model} we describe
the prescriptions we develop to model the formation of the ICL component. In
Section~\ref{sec:massfunction} we discuss how our prescriptions affect the
predicted galaxy stellar mass function, and in Section~\ref{sec:haloprop} we
discuss how the predicted fraction of ICL varies as a function of
halo properties. In Section~\ref{sec:formation}, we analyse when the bulk of
the ICL is formed, and which galaxies provide the largest contribution. We then
study the correlation between the ICL and the properties of the corresponding
BCGs in Section~\ref{sec:bcgprop}, and analyse the metal content of the ICL in
Section~\ref{sec:metallicity}. Finally, we discuss our results and give our
conclusions in Section~\ref{sec:discussion}.

\section[]{N-body simulations}
\label{sec:sim}

In this study we use collisionless simulations of galaxy clusters, generated
using the `zoom' technique \citep*[][ see also
  \citealt{Katz_and_White_1993}]{Tormen_etal_1997}: a target cluster is
selected from a cosmological simulation and all its particles, as well as
those in its immediate surroundings, are traced back to their Lagrangian region
and replaced with a larger number of lower mass particles. Outside this
high-resolution region, particles of increasing mass are displaced on a
spherical grid. All particles are then perturbed using the same fluctuation
field used in the parent cosmological simulations, but now extended to smaller
scales. The method allows the computational effort to be concentrated on the
cluster of interest, while maintaining a faithful representation of the large
scale density and velocity. 

Below, we use 27 high-resolution numerical simulations of regions around galaxy
clusters, carried out assuming the following cosmological parameters:
$\Omega_m=0.24$ for the matter density parameter, $\Omega_{\rm bar}=0.04$ for
the contribution of baryons, $H_0=72\,{\rm km\,s^{-1}Mpc^{-1}}$ for the
present-day Hubble constant, $n_s=0.96$ for the primordial spectral index, and
$\sigma_8=0.8$ for the normalization of the power spectrum. The latter is
expressed as the r.m.s. fluctuation level at $z=0$, within a top-hat sphere of
$8\,\hm$Mpc radius. For all simulations the mass of each Dark Matter particle 
in the high resolution region of
$10^8\,\hm {\rm M}_{\odot}$, and a Plummer-equivalent softening length is fixed to
$\epsilon=2.3 \hm$~kpc in physical units at $z<2$, and in comoving units at
higher redshift.

Simulation data have been stored at 93 output times, between $z=60$ and
$z=0$. Dark matter haloes have been identified using a standard
friends-of-friends (FOF) algorithm, with a linking length of 0.16 in units of
the mean inter-particle separation in the high-resolution region. The algorithm
{\small SUBFIND} \citep{Springel_etal_2001} has then been used to decompose
each FOF group into a set of disjoint substructures, identified as locally
overdense regions in the density field of the background halo. As in previous
work, only substructures that retain at least 20 bound particles after a
gravitational unbinding procedure are considered to be genuine substructures.
Finally, merger histories have been constructed for all self-bound structures
in our simulations, using the same post-processing algorithm that has been
employed for the Millennium Simulation \citep{springel3}. For more details on
the simulations, as well as on their post-processing, we refer the reader to
\citet{myself}. For our analysis, we use a sample of 341 haloes extracted from
the high resolution regions of
these simulations, with mass larger than $10^{13} \hm M_{\odot}$. In
Table~\ref{tab:tab0}, we give the number of haloes in different mass ranges.

\begin{table}
\caption{The sample of haloes used in this study, split in five subsamples
  according to their mass. The first column indicates the mass range, and the
  second column gives the number of haloes in each subsample.}
\begin{center}
\begin{tabular}{lllll}
\hline
Halo mass range & Number of haloes \\
\hline

$ \geq 10^{15}\,\hm {\rm M}_{\odot}$ & 13 \\
$[$5-10$]\times10^{14}\,\hm {\rm M}_{\odot}$ &  15 \\ 

$[$1-5$]\times10^{14}\,\hm {\rm M}_{\odot}$ & 25 \\ 

$[$5-10$]\times10^{13}\,\hm {\rm M}_{\odot}$ & 29 \\ 

$[$1-5$]\times10^{13}\,\hm {\rm M}_{\odot}$ & 259 \\
\hline
\end{tabular}
\end{center}
\label{tab:tab0}
\end{table}

\section[]{Semi-analytic models for the formation of the ICL}
\label{sec:model}

In this study, we use the semi-analytic model presented in DLB07, but we update
it in order to include three different prescriptions for modelling the
formation of the ICL component. These prescriptions are described in detail in
the following. For readers who are not familiar with the terminology used
within our model, we recall that we consider three different types of galaxies: 

\begin{itemize}
 \item Type 0: these are central galaxies, defined as those located at the
   centre of the main halo\footnote{This is the most massive subhalo of a FOF,
     and typically contains about 90 per cent of its total mass.} of each FOF
   group;
 \item Type 1: these are satellite galaxies associated with a distinct dark
   matter substructure. A Type 0 galaxy becomes Type 1 once its parent halo is
   accreted onto a more massive system;
 \item Type 2: also called \emph{orphan} galaxies, these are satellites whose
   parent substructures have been stripped below the resolution limit of the
   simulation. Our reference model assumes that, when this happens, the baryonic component
   is unaffected and the corresponding galaxy survives for a residual time
   before merging with the corresponding central galaxy.
   The position of an orphan galaxy is traced by following the position of the 
   particle that was the most bound particle of the parent substructure at 
   the last time it was identified.
\end{itemize}

The residual merger time assigned to each galaxy that becomes Type 2 is
estimated using the following implementation of the Chandrasekhar dynamical
friction formula:
\begin{equation}\label{dff}
  \tau_{merge} = f_{fudge}\frac{1.17}{\ln
    \Lambda}\frac{D^2}{R^2_{vir}}\frac{M_{par}}{M_{sat}}\tau_{dyn}
\end{equation}
where $D$ is the distance between the satellite and the centre of its parent
FOF, $R_{vir}$ is the virial radius of the parent halo, $M_{sat}$ the sum
of the dark and baryonic mass of the satellite, $M_{par}$ the (dark
matter) mass of the accreting halo, $\tau_{dyn} = R_{vir}/V_{vir}$ is the
dynamical time of the parent halo, and $\Lambda = 1+M_{par}/M_{sat}$ is the
Coulomb logarithm. 
All quantities entering equation \ref{dff} are evaluated at the last time the 
substructure hosting the satellite galaxy is identified, before falling 
below the mass limit for substructure identification. 
As in DLB07, we have assumed $f_{fudge} = 2$, which is in
better agreement with recent numerical work indicating that the classical
dynamical friction formulation tends to under-estimate the merging times
measured from simulations \citep{bk, Jiang_etal_2008}.

The reference model does not include a prescription for the formation of the
ICL. Below, we describe three different models \footnote{That are ''either/or'' 
prescriptions.} that we have implemented to
account for the ICL component. We assume that it is formed through two
different channels: (i) stellar stripping from satellite galaxies and (ii)
relaxation processes that take place during mergers \footnote{That we couple
with each of the three stellar stripping prescriptions.} and may unbind some
fraction of the stellar component of the merging galaxies. In the following, we
describe in detail each of our prescriptions.

\subsection[]{Disruption Model}

This model is equivalent to that proposed by \cite{qiguo}, and assumes that the
stellar component of satellite galaxies is affected by tidal forces only
after their parent substructures have been stripped below the resolution of the
simulation (i.e. the galaxies are Type 2). We assume that each satellite galaxy
orbits in a singular isothermal potential,

\begin{equation}
 \phi(R) = V_{vir}^2 \ln R,
\end{equation}
and assume the conservation of energy and angular momentum along the orbit to
estimate its pericentric distance:

\begin{equation}
\left(\frac{R}{R_{peri}}\right)^2 = \frac{\ln R/R_{peri} + \frac{1}{2}
  (V/V_{vir})^2}{\frac{1}{2} (V_t/V_{vir})^2}  .
\end{equation}
In the equation above, $R$ is the distance of the satellite from the halo
centre, and $V$ and $V_t$ are the velocity of the satellite with respect to
the halo centre and its tangential part, respectively.  Following
\citet{qiguo}, we compare the main halo density at pericentre with the average
baryon mass density (i.e. the sum of cold gas mass and stellar mass) of the
satellite within its half mass radius. Then, if the following condition is
verified:

\begin{equation}\label{eq2}
\frac{M_{DM, halo}(R_{peri})}{R_{peri}^3} = \rho_{DM,halo} > \rho_{sat} =
\frac{M_{sat}}{R_{half}^3} \, ,
\end{equation}
we assume the satellite galaxy to be disrupted and its stars to be assigned to
the ICL component of the central galaxy. In the equation above, we approximate
$R_{half}$ by the mass weighted average of the half mass radius of the disk and
the half mass radius of the bulge, and $M_{sat}$ is the baryonic mass 
(cold gas plus stellar mass). The cold \footnote{In our model, no hot
  component is associated with satellite galaxies.} gas mass that is associated
with the disrupted satellite is added to the hot component of the central
galaxy. When a central Type 0 galaxy is accreted onto a larger system and
becomes a Type 1 satellite, it carries its ICL component until its parent
substructure is stripped below the resolution limit of the simulation. At this
point, its ICL is added to that of the new central galaxy.

In a recent paper, \cite{alvaro} discuss the limits of this implementation with
respect to results from controlled numerical simulations of the evolution of
disk galaxies within a group environment. The model discussed above is applied
only after the galaxy's dark matter subhalo has been completely disrupted, but
the simulations by Villalobos et al. show that the stellar component of the
satellite galaxy can be significantly affected by tidal forces when its parent
subhalo is still present. In addition, this model assumes that the galaxy is
completely destroyed when equation (\ref{eq2}) is satisfied, but the
simulations mentioned above show that galaxies can survive for a relatively
long time (depending on their initial orbit) after they start feeling the tidal
forces exerted by the cluster potential. 

Predictions from this model are affected by numerical resolution. We carried out
a convergence test by using a set of low-resolution simulations with the same initial 
conditions of the high-resolution set used in this paper, but with the dark matter 
particle mass one order of magnitude larger than the one adopted in the high resolution set
and with gravitational softening increased accordingly by a factor $10^{1/3}$. 
We find that, on average, the ICL fraction is $\sim 30$ per cent higher in the 
low-resolution set ($\sim 20$ per cent in group-like haloes with mass $\sim 10^{13} M_{\odot} \hm$
and $\sim 50$ in the most massive haloes considered in our study, as shown in the 
right panel of Figure \ref{fig:ICLconv} in Appendix \ref{sec:numconv}). This is due to the fact 
that, decreasing the resolution, a larger fraction of satellite galaxies  
are classified as Type 2 and are subject to our stripping model.
While the lack of numerical convergence does not affect the qualitative conclusions 
of our analysis, we point out that the amount of ICL measured in our simulations should 
be regarded as an upper limit.

\subsection[]{Tidal Radius Model}

In this prescription, we allow each satellite galaxy to lose mass in a continuous
fashion, before merging or being totally destroyed. Assuming that the stellar
density distribution of each satellite can be approximated by a spherically
symmetric isothermal profile, we can estimate the \emph{tidal radius} by means
of the equation:

\begin{equation}
  R_{t} = \left(\frac{M_{sat}}{3 \cdot M_{DM,halo}}\right)^{1/3} \cdot D
\end{equation} 
\citep{binney}. In the above equation, $M_{sat}$ is the satellite mass (stellar
mass + cold gas mass), $M_{DM,halo}$ is the dark matter mass of the parent
halo, and $D$ the satellite distance from the halo centre.  

In our model, a galaxy is a two-component system with a spheroidal component
(the bulge), and a disk component.  If $R_t$ is smaller than the bulge radius,
we assume the satellite to be completely disrupted and its stellar and cold
mass to be added to the ICL and hot component of the central galaxy,
respectively. If $R_t$ is larger than the bulge radius but smaller than the
disk radius, we assume that the mass in the shell $R_t -R_{sat}$ is stripped
and added to the ICL component of the central galaxy. A proportional fraction
of the cold gas in the satellite galaxy is moved to the hot component of the
central galaxy. We assume an exponential profile for the disk, and $R_{sat} =
10 \cdot R_{sl}$, where $R_{sl}$ is the disk scale length ($R_{sat}$ thus
contains 99.9 per cent of the disk stellar mass).  After a stripping episode,
the disk scale length is updated to one tenth of the tidal radius.

This prescription is applied to both kinds of satellite galaxies.
For Type 1 galaxies, we derive the tidal radius including the dark 
matter component in $M_{sat}$, and we impose that stellar stripping can take 
place only if the following condition is verified:

\begin{equation}\label{eqn:eq_radii}
 R^{DM}_{half} < R^{Disk}_{half} \, ,
\end{equation}
where $R^{DM}_{half}$ is the half-mass radius of the parent subhalo, and
$R^{Disk}_{half}$ the half-mass radius of the galaxy's disk, that is $1.68\cdot
R_{sl}$ for an exponential profile. When a Type 1 satellite is affected by
stellar stripping, the associated ICL component is added to that of the
corresponding central galaxy. 

As for the Disruption model, predictions from the Tidal Radius model are affected 
by numerical resolution. We carried out the same convergence test used for the 
Disruption model. Again, we find for the low resolution set a larger amount of ICL (by 
about 40 per cent, almost independent of halo mass, as shown in the left panel of Figure 
\ref{fig:ICLconv} in Appendix \ref{sec:numconv}). In this model, Type 2 
galaxies are the dominant contributors to the ICL and, as for the Disruption model, 
the larger number of these satellites in the low-resolution set translates in a larger 
number of galaxies eligible for tidal stripping.

\subsection[]{Continuous Stripping Model}
\label{sec:modc}

This model is calibrated on recent numerical simulations by \citet{alvaro}.
These authors have carried out a suite of numerical simulations aimed to study
the evolution of a disk galaxy within the global tidal field of a group
environment (halo mass of about $10^{13}\,{\rm M}_{\odot}$). In the
simulations, both the disk galaxy and the group are modelled as multi-component
systems composed of dark matter and stars.
The evolution of the disk galaxy is followed after it crosses the group virial radius,  
with initial velocity components consistent with infalling substructure from cosmological 
simulations (see \citealt{benson}).
The simulations cover a broad
parameter space and allow the galaxy-group interaction to be studied as a
function of orbital eccentricity, disk inclination, and galaxy-to-group mass
ratio. We refer to the original paper for a more detailed description of the
simulations set-up, and of the results.

Analysing the outputs of these simulations (Villalobos et al, in preparation),
we have derived a fitting formula that describes the evolution of the stellar
mass lost by a satellite galaxy as a function of quantities estimated at the
time of accretion (i.e. at the time the galaxy crosses the virial radius of the
group). Our fitting formula reads as follow:

\begin{equation}
  M^*_{lost} =
  M^*_{accr}\exp\left[\left(\frac{-16}{1-\eta}\right)\left(\frac{M_{sub}}{M_{par}}\right)^{\frac{1}{2}}\left(1-\frac{t}{t_{merg}} 
    \right) \right]
\label{eq:modc}
\end{equation}
where $M^*_{accr}$ is the stellar mass at the time of accretion, $\eta$ is the
circularity of the orbit, $M_{sub}$ and $M_{par}$ are the subhalo and parent
halo dark matter masses respectively, and $t_{merg}$ is the residual merger
time of the satellite galaxies. We approximate the accretion time as the last
time the galaxy was a central galaxy (a Type 0), and compute the circularity 
using the following equation:

\begin{equation}\label{eta}
  \eta = V_{\theta}\sqrt{\frac{2f-V^2_r -V^2_{\theta}}{2f-1}},
\end{equation}
where $V_r$ and $V_{\theta}$ are the radial and tangential velocities of the
accreted subhalo, and $f=1+M_{sub}/M_{par}$. For each accreted galaxy, $V_r$
and $V_{\theta}$ are extracted randomly from the distributions measured by
\citet{benson} from numerical simulations.

Since Eq.~\ref{eq:modc} is estimated at the time of accretion, we cannot use
the merger time prescription that is adopted in the reference model, where a
residual merger time is assigned only at the time the substructure is stripped
below the resolution of the simulation. To estimate merger times at the time of
accretion, we use the fitting formula by \cite{bk} (hereafter BK08), that has
been calibrated at the time the satellite galaxy crosses the virial radius of
the accreting system, and is therefore consistent with our Eq.~\ref{eq:modc}.
In a few cases, it happens that the merger time is elapsed when the
satellite galaxy is still a Type 1. In this case, we do not allow the galaxy
to merge before it becomes a Type 2.

\citet{gab2010} compared the merger times predicted using this formula with no
orbital dependency and for circular orbits with those provided by Equation
\ref{dff} used in the reference model, and found a relatively good
agreement. We find that, once the orbital dependency is accounted for, the BK08
fitting formula predicts merger times that are on average shorter than those
used in our reference model by a factor of $\sim 3$ for mass-ratios
$M_{sub}/M_{par} > 0.025$. Therefore, in our continuous stripping model, mergers
are on average shorter than in the disruption and tidal radius models. 

To implement the model described in this section for the formation of the ICL,
we use Eq.~\ref{eq:modc} to compute how much stellar mass has to be removed
from each satellite galaxy at each time-step. The stripped stars are then added
to the ICL component of the corresponding central galaxy, and a proportional
fraction of the cold gas in the satellite is moved into the hot component
associated with the central galaxy. If the stripped satellite is a Type 1, and
it carries an ICL component, this is removed at the first episode of stripping
and added to the ICL component of the central galaxy.

It is worth stressing that Eq.~\ref{eq:modc} is valid for disk galaxies that
are accreted on a system with velocity dispersion typical of a galaxy group,
and that we are extrapolating the validity of this equation to a wider halo
mass ranges. 

\subsection[]{Merger channel for the formation of the ICL}
\label{sec:mergers}

\citet{giuseppe} argue that the bulk of the ICL is not due to tidal stripping
of satellite galaxies (that in their simulations accounts for no more than 5-10
per cent of the total diffuse stellar component), but to relaxation processes
taking place during the mergers that characterize the build-up of the
central dominant galaxy.

In our reference model, if two galaxies merge, the stellar mass of the merging
satellite is added to the stellar (bulge) mass of the central galaxy. Based on
the findings described above, we add a `merger channel' to the formation of the
ICL by simply assuming that, when two galaxies merge, 20 per cent of the
satellite stellar mass gets unbound and is added to the ICL of the
corresponding central galaxy. We have verified that this simple prescriptions
reproduces approximately the results of the numerical simulations by
\cite{alvaro}, though in reality the fraction of stars that is unbound should
depend on the orbital circularity (Villalobos et al., in preparation).
We have also verified that assuming that a larger fraction of the satellite 
stellar mass gets unbound, obviously leads to higher ICL fractions. In particular, 
assuming that 50 per cent of the stellar mass of the satellite is unbound, almost 
doubles the ICL fractions predicted. Assuming an even higher fraction does not 
affect further the ICL fraction because the effect of having more stellar mass 
unbound is balanced by the fact that merging galaxies get significantly less massive.

A similar prescription was adopted in \citet{pigi} who showed that this has
important consequences on the assembly history of the most massive galaxies,
and in \citet{somerville} as one possible channel for the formation of the ICL
in the context of a hierarchical galaxy formation model. In order to test the
influence of this channel on our results, in the following we will present
results with this channel both on and off.

\subsection[]{Modelling the bulge and disk sizes}

Our reference model (DLB07) does not include prescriptions to model the bulge
size, that we use in our stellar stripping models. To overcome this limitation,
we have updated the reference model including the prescriptions for bulge and
disk growth described in \cite{qiguo}.

Both the gaseous and stellar components of the disk are assumed to follow an
exponential profile.  Assuming a flat circular velocity curve, the
scale-lengths of these two components can be written as:
\begin{equation}
  R^g_{sl}=\frac{J_{gas}/M_{gas}}{2V_{max}}, \hspace{1cm}
  R^*_{sl}=\frac{J_{*}/M_{*,disk}}{2V_{max}},
\end{equation}
where $J_{gas}$ and $J_*$ are the angular momenta of the gas and stars,
$M_{gas}$ and $M_{*,disk}$ are the gas and stellar mass of the disk, and
$V_{max}$ is the maximum circular velocity of the dark halo associated with the
galaxy. Following \citet{qiguo}, we assume that the change in the angular
momentum of the gas disk during a timestep can be expressed as the sum of the
angular momentum changes due to addition of gas by cooling, accretion from
minor mergers, and gas removal through star formation. The latter causes a
change in the angular momentum of the stellar disk. We refer to the original
paper by Guo et al. for full details. We have verified that switching back to
the simple model for disk sizes of \citet{mo98} used in our reference model,
does not affect significantly the results discussed in the
following\footnote{Note that the disk size enters in the calculation of the
  star formation rate. Therefore, a change in the disk size model could, in
  principle, affect significantly model results.}.

Bulges grow through two different channels: mergers (both major and minor) and
disk instability. Following \citet{qiguo}, we estimate the change in size due
to a merger using energy conservation and the virial theorem:
\begin{equation}
  \frac{G M_{new,b}^2}{R_{new,b}} = \frac{G M_1^2}{R_1} + 
                                       \frac{G M_2^2}{R_2} + 
                                       \frac{M_1 M_2}{R_1+R_2} 
\end{equation}
The same approach is adopted in case of disk instability, simply replacing
$R_1$ and $M_1$ with the size and mass of the existing bulge, and $R_2$ and
$M_2$ with the size and mass of the stars that are transferred from the disk to
the bulge so as to keep the disk marginally stable. $R_2$ is determined
assuming that the mass is transferred from the inner part of the disk, with the
newly formed bulge occupying this region. Again, we refer to the original paper
by \citet{qiguo} for full details on this implementation.

\section[]{The Galaxy Stellar Mass Function}
\label{sec:massfunction}

\begin{center}
\begin{figure*}
\includegraphics[scale=.47]{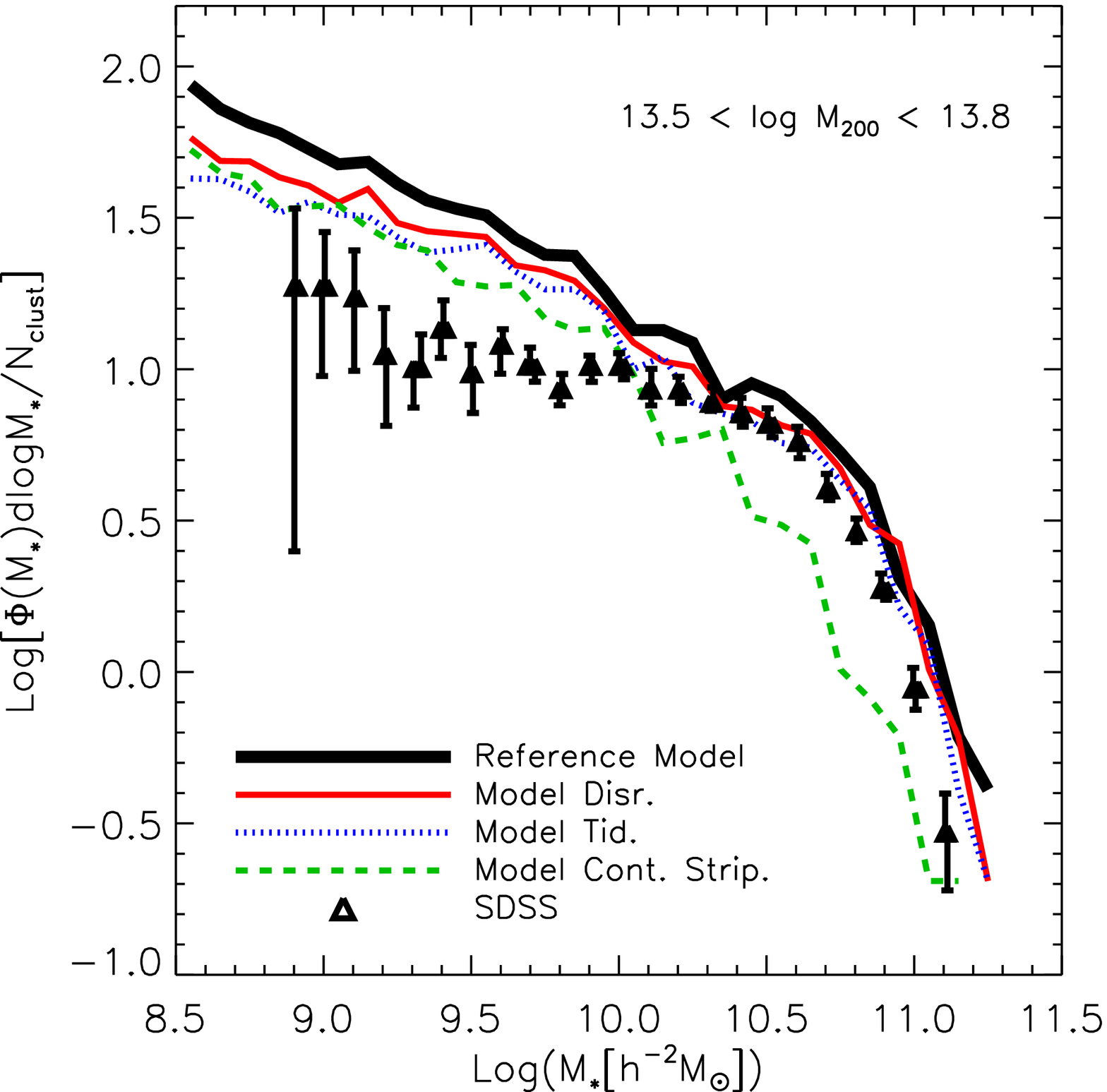} 
\includegraphics[scale=.47]{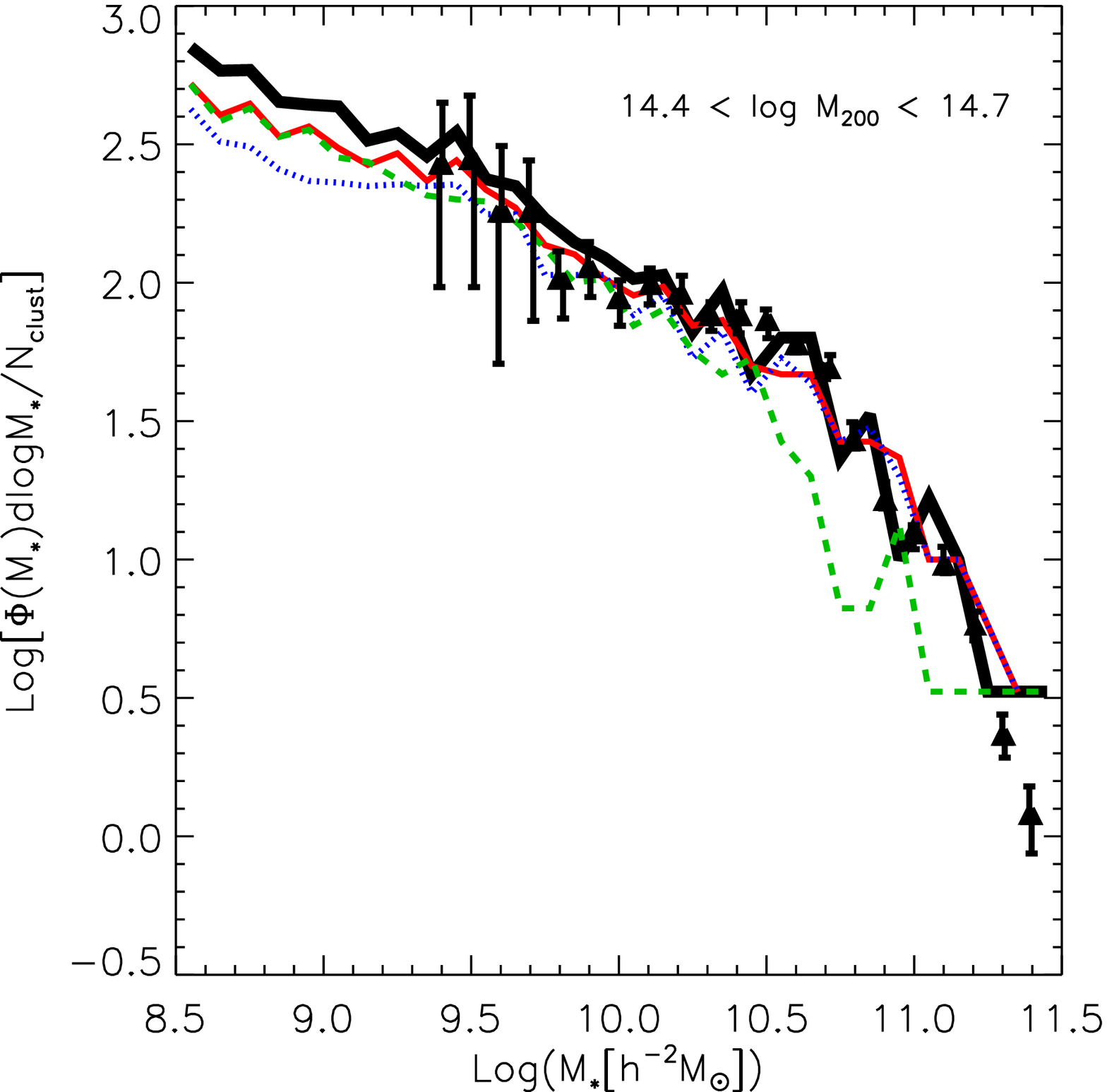} 
\caption{Left panel: The conditional stellar mass function of satellite
  galaxies in haloes in the mass range $13.5 < \log M_{200} [M_{\odot} \hm] <
  13.8$. Our simulations provide a total of 49 haloes in this mass
  range. Predictions from our reference model (DLB07) are shown by a solid
  black line, while predictions from our three models including disruption
  or/and stripping of satellite galaxies are shown by thinner lines of
  different style, as indicated by the legend. Symbols with error bars show
  observational measurements based on SDSS by \citet{liu10}. Right panel: as in
  the left panel, but for the 3 haloes from our simulations with $14.4 < \log M_{200}
  [M_{\odot} \hm] < 14.7$.}
\label{fig:MF}
\end{figure*}
\end{center}
Before focusing our discussion on the ICL component, it is interesting to
analyse how the proposed prescriptions affect one basic prediction of our
model, that is the galaxy stellar mass function. In Figure~\ref{fig:MF}, we
show the conditional stellar mass function of satellite galaxies in the 49
haloes from our simulations that fall in the mass range $13.5 < \log M_{200}
[M_{\odot} \hm] < 13.8$ in the left panel, and in the 3 haloes from our
simulations with $14.4 < \log M_{200} [M_{\odot} \hm] < 14.7$ in the right panel. We
show predictions from both our reference model (DLB07, shown as a solid black
line), and from the three models including the treatments for the stripping
and/or disruption of satellite galaxies discussed in Section~\ref{sec:model}
(lines of different style, see legend). Model predictions are compared with
observational measurements by \citet{liu10}. These are based on group 
catalogues constructed from the Sloan Digital Sky Survey (SDSS) Data Release 
4, using all galaxies with extinction corrected magnitude brighter than $r=18$ 
and in the redshift range $0.01 \leq z \leq 0.2$.
We do not show here predictions from the models with the merger
channel for the formation of the ICL switched on, as these do not deviate
significantly from the corresponding models with the merger channel off.

For the lower halo mass range considered, the reference model fails in
reproducing the observed stellar mass function, over-predicting the abundance
of galaxies with stellar mass below $\sim 10^{10}\,{\rm M}_{\odot}$. This
problem is somewhat alleviated when including a model for stellar stripping,
but not solved. In particular, our `disruption model' (model Disr. in the
figure and hereafter) does not significantly affect the abundance of the most
massive satellites, while reducing the number of their lower mass counterparts
(not by the amount required to bring model predictions in agreement with
observational results). This is expected as this model only acts on Type 2
galaxies, that dominate the low-mass end of the galaxy mass function.

Predictions from the `tidal radius model' (model Tid.) are not significantly
different from those of model Disr., while the `continuous stripping model'
(model Cont. Strip.) significantly under-predicts the abundance of the most
massive satellite galaxies. There are two possible explanations for this
behaviour: (i) the abundance of massive satellites is reduced because these are
significantly affected by our stripping model or (ii) these massive satellites
have disappeared because they have merged with the central galaxies of their
parent haloes. As discussed in Section~\ref{sec:modc}, our model Cont. Strip.
uses a different prescription for merger times with respect to that employed in
the reference model and in models Disr. and Tid. We find that, in this model,
merger times are on average shorter than in the other models, which is the
reason for the under-prediction of massive satellites shown in
Figure~\ref{fig:MF}. As we will discuss in the following, this also implies
that the stellar mass of the BCGs in model Cont. Strip. are on average larger
than those predicted by models Disr. and Tid.

For the higher halo mass range considered, the observed number density of
intermediate-to-low mass galaxies is higher, and all our models appear to be in
agreement with observational measurements. The agreement remains good also at
the massive end, with the exception of model Cont. Strip. that significantly
under-predicts the number density of massive galaxies. As explained above, this
is due to the shorter galaxy merger times used in this model.

We will show below that, in our models, the merger channel does not provide the
dominant contribution to the ICL formation so that this is mainly driven by
stripping and/or disruption of satellites. Therefore, the excess of
intermediate to low-mass galaxies for haloes in the lower mass range might
invalidate our predictions. However, as we will show below, the bulk of the ICL
originates from relatively massive galaxies so that this particular failure of
our models does not significantly affect our results.

\section[]{ICL fraction and dependency on halo properties}
\label{sec:haloprop}

\begin{center}
\begin{figure*}
\includegraphics[scale=.48]{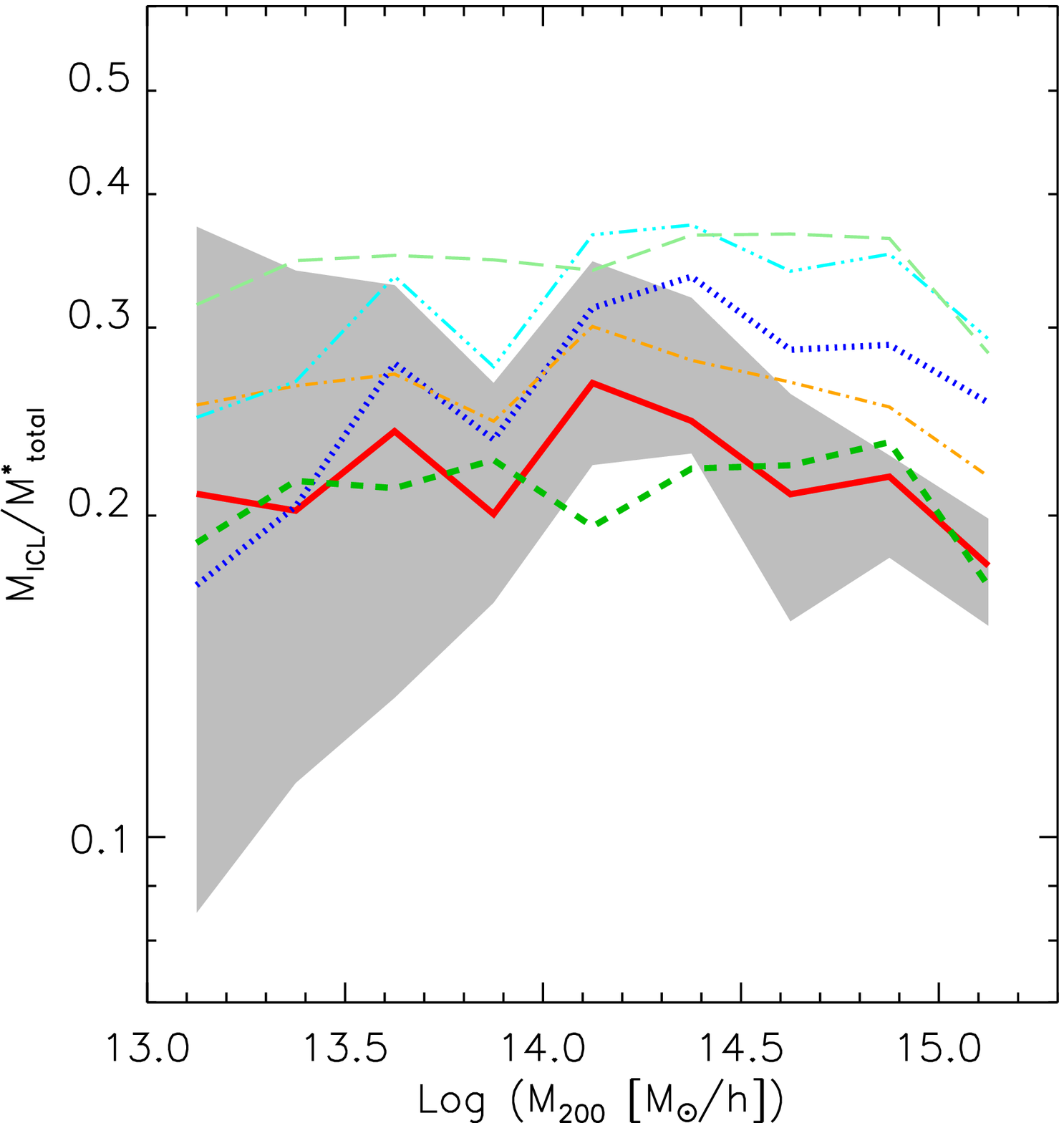}
\hspace{10pt}
\includegraphics[scale=.48]{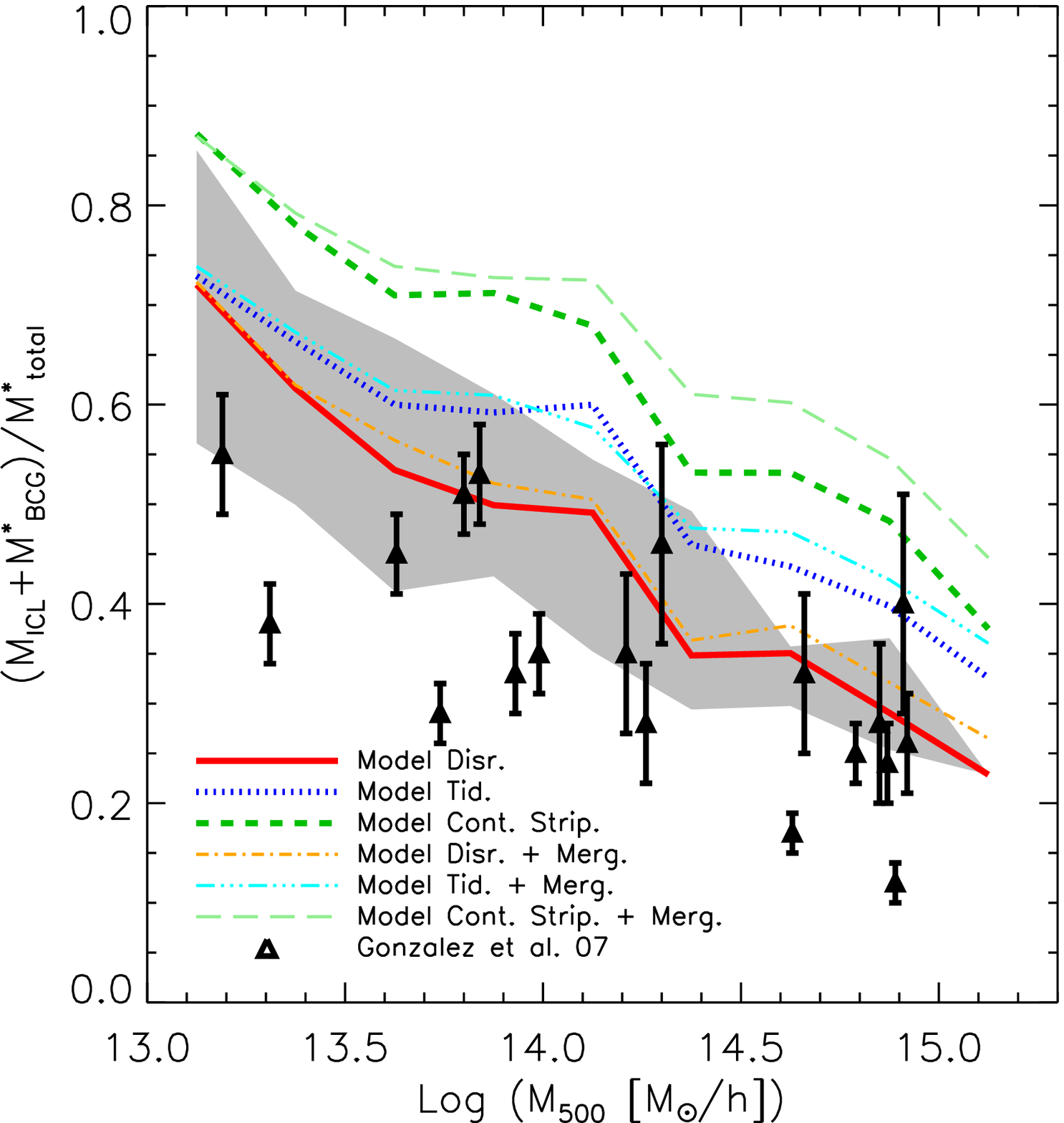}
\caption{Left panel: ICL fraction as a function of halo mass. Lines of
  different style and colour show median results from different models, as
  indicated in the legend. Right panel: ratio between the ICL component plus
  BCG stellar mass and the total stellar mass within $R_{500}$. Symbols with
  error bars show observational estimates by \citealt{gonzalez07}. In both
  panels, the gray shaded region shows the 20th and 80th percentiles of the
  distribution for model Disr. The other models have comparable scatter.}
\label{fig:ICLfrac}
\end{figure*}
\end{center}
We now turn our analysis to the ICL component, and start by analysing the
overall fraction of ICL predicted by our models, and how it depends on halo
properties. The left panel of Figure~\ref{fig:ICLfrac} shows the ICL fraction
as a function of halo mass. To measure the predicted fractions, we have
considered all galaxies within $R_{200}$ and with stellar mass larger than $M_*
= 10^{8.3} M_{\odot} $, that approximately corresponds to the resolution limit
of our simulations. Lines of different style and colour show the median
relations predicted by our different prescriptions, as indicated in the
legend. The grey region marks the 20th and 80th percentiles of the distribution
found for model Disr. (the other models exhibit a similar dispersion).

The predicted fraction of ICL varies between $\sim 20$ per cent for model Disr.
with the merger channel off, and $\sim 40$ per cent for model Cont. Strip. with
the merger channel on. A relatively large halo-to-halo variation is measured
for all models. For none of our models, we find a significant increase of the
ICL fraction with increasing halo mass (see discussion in Section
\ref{sec:intro}), at least over the range of $M_{200}$ shown.
In the figure, we have also considered the ICL associated with Type 1 galaxies
within $R_{200}$ of each halo. We stress, however, that their contribution is,
on average, smaller than 7 per cent of the total ICL associated with the
cluster. 

As expected, when including a merger channel for the formation of the ICL, its
fraction increases, by about $\sim 25$ per cent in models Disr. and Tid., and by
about $\sim 50$ per cent in model Cont. Strip. As mentioned in the previous
Section, this difference is due to the different dynamical friction formula
used in this model, which makes merger times significantly shorter than in the
other two models.  The amount of ICL that comes from the merger channel is not
negligible in the framework of our models and, as expected, increases in the
case of shorter merger times. Overall, predictions from our models agree well
with fractions of ICL quoted in the literature, i.e. $10-40$ per cent going
from groups to clusters (e.g. \citealt{feldmeier04,zibetti08,mcgee10,toledo11}).

As discussed in Section~\ref{sec:intro}, it is not an easy task to separate the
ICL from the stars that are bound to the BCG. To avoid these difficulties, and
possible biases introduced by the adoption of different criteria in the models
and in the observations, we consider the ratio
$(M_{ICL}+M_{BCG})/M^*_{total}$. The right panel of Figure~\ref{fig:ICLfrac}
shows how this fraction varies for the different models considered in this
study, and compares our model predictions with observational measurements by
\cite{gonzalez07}. 
To be consistent with these observational measurements, we
consider in this case all galaxies within $R_{500}$ and brighter than $m_I =
18$ (i.e. galaxies more massive than $\sim 10^{10} M_{\odot}$).
Considering the scatter in both our model predictions and in the
observational data, the Figure shows that model Disr. (as well as its variation
with the merger channel on) is in relatively good agreement with observational
data, though it tends to predict higher ratios for the lowest halo masses
considered. Model Tid. predicts a higher fraction of stars in ICL+BCG than
model Disr., and the median relation lies close to the upper envelope of the
observational data. Finally, model Cont. Strip. over-predicts the fraction of
stars in ICL+BCG over the entire mass range considered.

The merger channel does not affect the predicted trend as a function of halo
mass, but this channel increases on average the fraction of stars in ICL+BCG.
This is surprising: stars that are contributed to the ICL through the merger
channel would contribute to the BCG mass if the channel is off so that the
merger channel should not affect the sum of the ICL and BCG stellar masses.
The difference found is due to slight changes in the merger history of the
BCGs, due to variations in the merger times of satellite galaxies.  In fact,
satellite merger times become slightly longer than in a model that does not
include stellar stripping because satellite galaxies gain less stellar mass
through accretion as central galaxies (also the satellites that were accreted
on them were stripped). Because of the shorter merger times mentioned above,
model Cont. Strip. is affected more by this change.

\begin{center}
\begin{figure*}
\includegraphics[scale=.503]{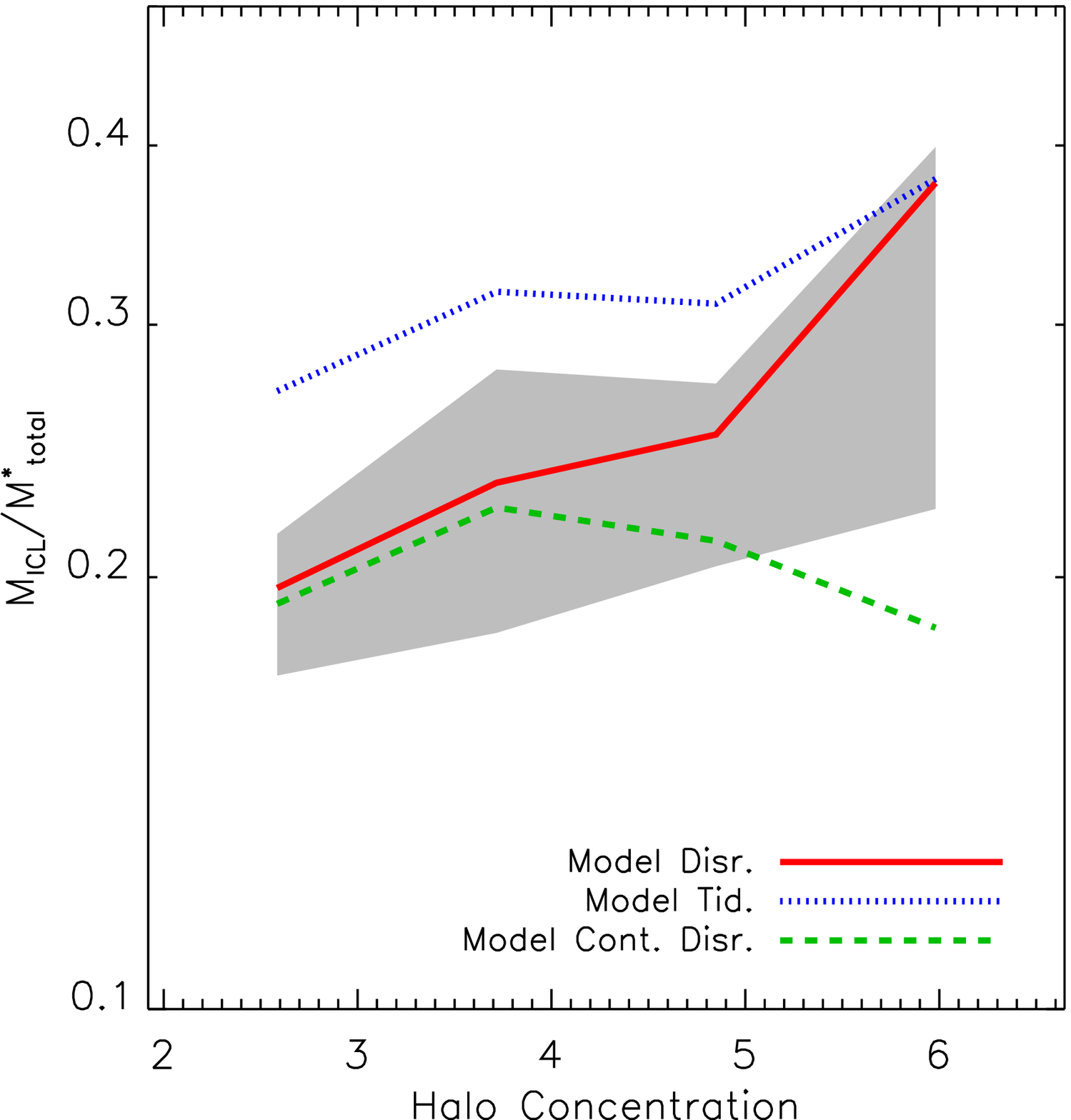}
  \hspace{10pt}
\includegraphics[scale=.50]{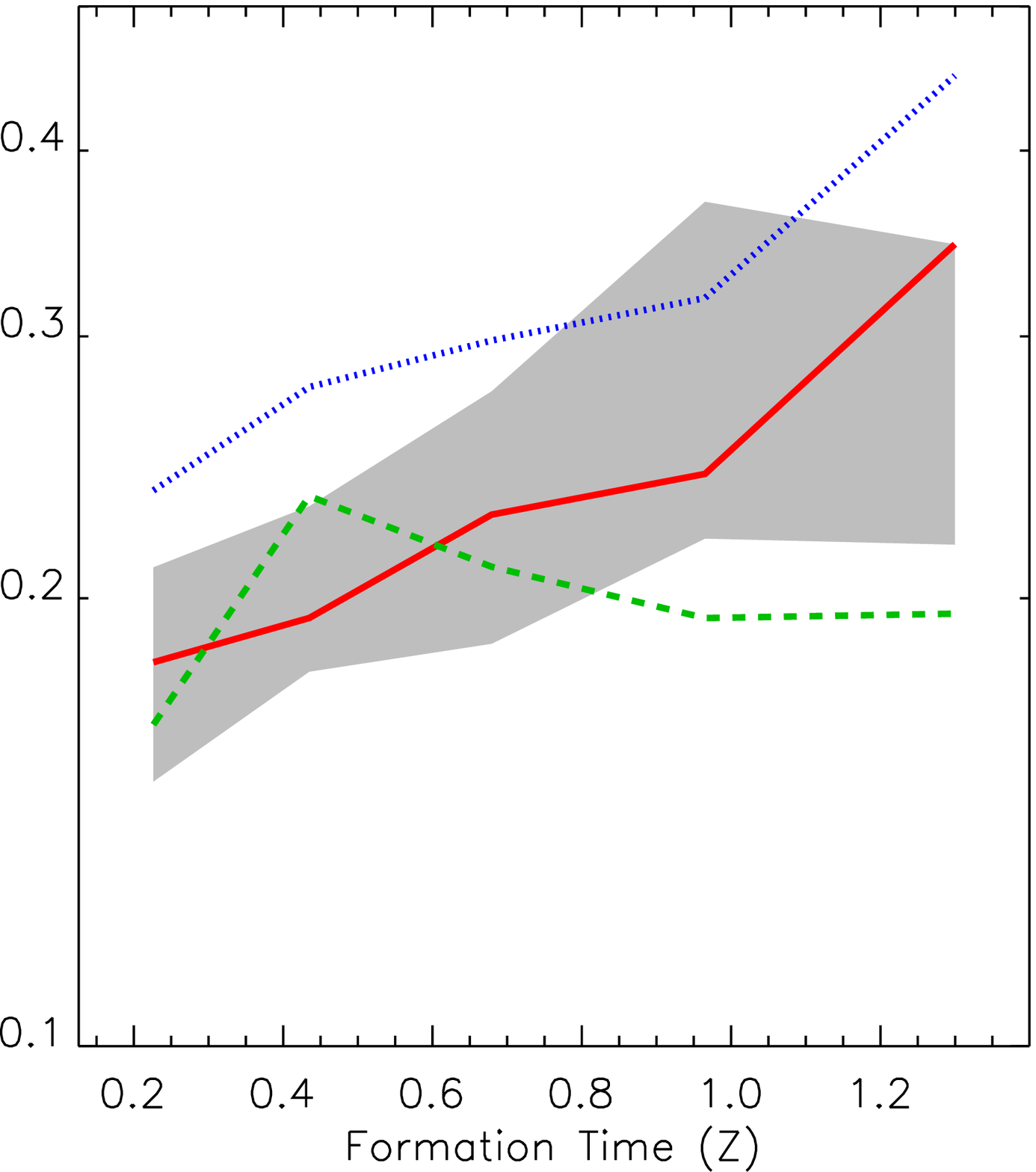}
\caption{Left Panel: Fraction of ICL as a function of halo concentration for
  the 53 haloes from our simulations with $M_{200}[M_{\odot}/h] >
  10^{14}$. Halo concentration is defined as $c_{200} = r_s /r_{200}$, where
  $r_s$ is the characteristic scale obtained by fitting the halo density
  profile to a NFW profile.  Right Panel: Fraction of ICL as a function of halo
  formation time, defined as the time when the main progenitor of the halo
  has acquired 50 per cent of its final mass.  Predictions from different
  models are shown by lines of different colour, as indicated in the
  legend. The gray shaded region shows the 20th and 80th percentile of the
  distribution found for model Disr.}
\label{fig:ICLconc}
\end{figure*}
\end{center}

As discussed above, the ICL fraction predicted by our models does not vary as a
function of the host halo mass but exhibits a relatively large halo-to-halo
scatter, particularly at the group mass scale. The natural expectation is that
this scatter is largely determined by a variety of mass accretion histories at
fixed halo mass. We can address this issue explicitly using our simulations. In
Figure~\ref{fig:ICLconc}, we show how the ICL fraction correlates with the halo
concentration in the left panel, and with the halo formation time in the right
panel. We note that these two halo properties are correlated \citep[see
  e.g.][]{giocoli12}. As usually done in the literature, we have defined the
formation time of the halo as the time when its main progenitor has acquired
half of its final mass. The concentration has been computed by fitting the
density profile of the simulated haloes with a NFW profile \citep*{nfw}. To
remove the known correlations between halo mass and concentration/formation
time \citep[][]{bullock01,Neto_etal_2007,power12}, we consider in
Figure~\ref{fig:ICLconc} only the 53 haloes from our simulations with
$M_{200}[M_{\odot}/h] > 10^{14}$. As in previous figures, we show the
dispersion (20th and 80th percentiles of the distribution) only for model
Disr. The other two models exhibit a similar scatter.

The figure shows that, for the halo mass range considered, the ICL fraction
increases with increasing concentration/formation time for models Disr. and Tid
while remaining approximately constant for model Cont. Strip. Therefore, for
models Disr. and Tid., large part of the scatter seen in
Figure~\ref{fig:ICLfrac} for haloes in the mass range considered can be
explained by a range of dynamical histories of haloes. Haloes that `formed' earlier
(those were also more concentrated) had more time to strip stars from
their satellite galaxies or accumulate ICL through accretion of smaller
systems, and therefore end up with a larger fraction of ICL. For model Cont.
Strip., no clear trend as a function of either halo concentration or halo
formation time is found. This happens because, contrary to the other two
models, the way we model stellar stripping in the Cont. Strip. model does not
introduce any dependence on halo concentration. In this model, 
the amount of stellar mass stripped from satellite galaxies depends only on
properties computed at the time of accretion (i.e. mass ratio, circularity of
the orbit). In contrast, in models Disr. and Tid., the stripping efficiency is
computed considering the instantaneous position of satellite galaxies. 
Dynamical friction rapidly drags more massive satellites (those contributing
more to the ICL) towards the centre, where stripping becomes more efficient.
Our results show that equation \ref{eq:modc} is not able to capture this 
variation, although it is by construction included in the simulations used to 
calibrate our model.

For lower mass haloes, there is no clear correlation between ICL fraction and
the two halo properties considered. This is in part due to the fact that, as we
will see in the next section, the bulk of the ICL forms very late - later than
the typical formation time of low-mass haloes. In this low-mass range, large
part of the scatter in the ICL fraction is driven by the fact that these haloes
typically contain relatively few massive galaxies, that are those contributing
the bulk of the ICL (see next section). So it is the scatter in the accretion
of single massive galaxies that drives the relatively large dispersion seen in
Figure~\ref{fig:ICLfrac} for haloes with mass smaller than $10^{14}\,{\rm
  M}_{\odot}$. We have explicitly verified this by considering the 20 per cent
haloes in this mass range with the highest and lowest ICL fractions. We find
that for those haloes that have larger ICL components, this has been
contributed by a few relatively massive satellite galaxies (stellar mass larger
than $10^{10}\,{\rm M}_{\odot}$). In contrast, the ICL in haloes with lower ICL
fractions was contributed by less massive satellite galaxies.

\begin{center}
\begin{figure}
\includegraphics[scale=.43]{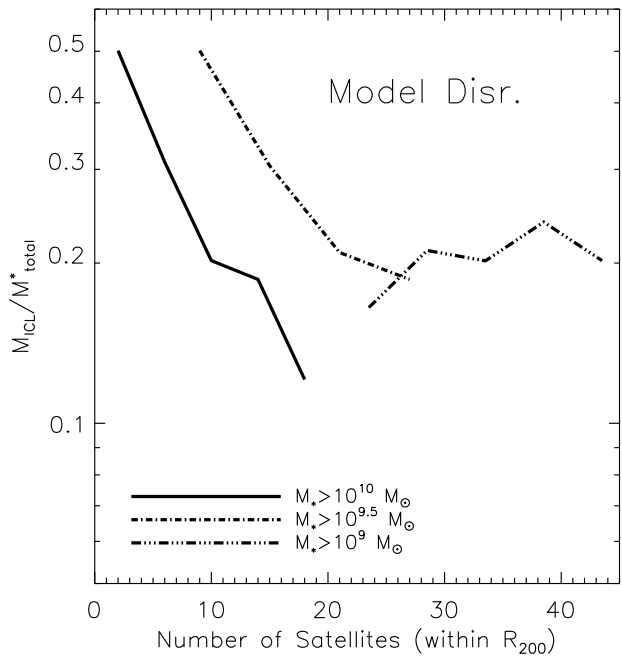} 
\caption{Fraction of ICL as a function of the number of satellites within
  $R_{200}$, with mass larger than the thresholds indicated in the legend. The
  figure refers to haloes in the mass range $10^{13.4} < M_{200} [M_{\odot}
    \hm] < 10^{13.6}$, and to the model Disr.}
\label{fig:ICLnumsat}
\end{figure}
\end{center}

In a recent work, \citet{purcell07} study how the fraction of ICL varies over a
wide range of halo masses (from that of spiral galaxies like our Milky Way to
that of massive galaxy clusters) using an analytic model for subhalo infall and
evolution, and empirical constraints to assign a stellar mass to each accreted
subhalo. In their model, the stellar mass associated with a subhalo is assumed
to be added to the diffuse component when a certain fraction of the subhalo
dark matter mass has been stripped by tidal interaction with the parent
halo. In particular, their fiducial model assumes that disruption of the
stellar component starts when 20 per cent of the subhalo mass remains
bound. Over the range of halo masses sampled in our study, \citet{purcell07}
predict a weak increase of the ICL fraction, from $\sim 20$ per cent for haloes
of mass $\sim 10^{13}\,{\rm M}_{\odot}$ to about 30 per cent for massive galaxy
clusters of mass $\sim 10^{15}\,{\rm M}_{\odot}$. One specific prediction of
their model is that the fraction of ICL correlates strongly with the number of
surviving satellite galaxies. As they explain, this correlation arises from the
fact that haloes that acquired their mass more recently had relatively less
time to disrupt the subhaloes they host, and therefore have less ICL.

We analyse the same correlation within our models in
Figure~\ref{fig:ICLnumsat}. This shows the ICL fraction as a function of the
number of satellite galaxies within $R_{200}$ for haloes in the mass range
$10^{13.4} < M_{200} [M_{\odot} \hm] < 10^{13.6}$, and for model Disr.. Lines of
different style correspond to different mass cuts, as indicated in the
legend. The figure shows that, when considering all galaxies with stellar mass
larger than $10^9\,{\rm M}_{\odot}$, no significant trend is found between the
ICL fraction and the corresponding number of surviving satellite galaxies (this
number ranges between $\sim 20$ and $\sim 45$ for this stellar mass cut). When
increasing the stellar mass cut, the number of surviving satellites decreases
(as expected), and a trend appears in the sense that larger ICL fractions are
measured for lower numbers of surviving satellites. Similar trends are found
for model Tid., while for model Cont. Strip. no trend is found between the
number of surviving satellites and the ICL fraction, for any mass cut
used.

\section[]{Formation of the ICL}
\label{sec:formation}

In this section, we take advantage of our models to analyse when the bulk of
the ICL forms, which galaxies provide the largest contribution to it, what is
the fraction of ICL that has been accreted from other haloes during the
hierarchical growth of clusters, and what fraction is instead contributed from
the merger channel. 

\begin{figure}
\begin{center}
\includegraphics[scale=.45]{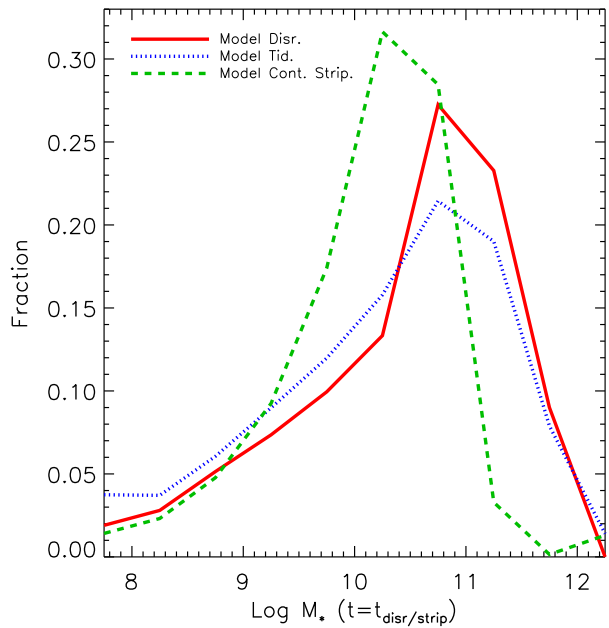} 
\caption{Fraction of the ICL component as a function of the galaxy stellar mass
  that contributed to it. In the case of model Disr., the galaxy stellar mass
  on the x-axis refers to the mass that the satellite has before its
  disruption. For model Cont. Strip., it corresponds to the satellite mass
  before stripping takes place. For model Tid., both cases can occur.}
\label{fig:ICLstarmass}
\end{center}
\end{figure}

In Figure~\ref{fig:ICLstarmass}, we show the contribution to the ICL from
galaxies with different stellar mass. For model Disr., the galaxy mass on the
x-axis corresponds to that of the satellite right before its disruption, while
for model Cont. Strip. it corresponds to the satellite mass before stripping
takes place. For model Tid., both cases can occur. The distributions shown in
Figure~\ref{fig:ICLstarmass} represent the average obtained considering all
haloes in our simulated sample. When considering haloes in different mass bins,
the distributions are similar but, as expected, they shift towards lower
stellar masses for lower mass haloes. 

The figure shows that the bulk of the ICL comes from galaxies with stellar
masses $\sim 10^{11}\,{\rm M}_{\odot}$ for models Disr. and Tid., and $\sim
10^{10.3}\,{\rm M}_{\odot}$ for model Cont. Strip. In particular, we find that
for models Disr. and Tid., about 26 per cent of the ICL is contributed by
galaxies with stellar mass in the range $10^{10.75}-10^{11.25}
M_{\odot}$. About 68 per cent comes from satellites more massive than
$10^{10.5} M_{\odot}$, while dwarf galaxies contribute very little. For model
Cont. Strip., almost all the ICL mass ($\sim 90$ per cent of it) comes from
satellites with mass in the range $10^{9}-10^{11} M_{\odot}$.  The merger
channel does not affect significantly the distributions shown.

The result discussed above can be easily understood in terms of dynamical
friction: the most massive satellites decay through dynamical friction to the
inner regions of the halo on shorter time-scales than their lower mass
counterparts. Tidal forces are stronger closer to the halo centre, so that the
contribution to the ICL from stripping and/or disruption of massive galaxies is
more significant than that from low mass satellites. The latter tend to spend
larger fractions of their lifetimes at the outskirts of their parent halo,
where tidal stripping is weaker.

The differences between predictions from models Disr. and Tid. and those from
model Cont. Strip. are due to a combination of different effects. On one side,
model Cont. Strip. uses a different merger time prescription that leads to
significantly shorter merger times than in models Disr. and Tid. This is
particularly important for the most massive satellites that have the shortest
merger times. In addition, while in models Disr. and Tid. satellite galaxies
can be completely destroyed, stripping takes place in a more continuous and
smooth fashion in model Cont. Strip. These two effects combine so that the
largest contribution to the ICL in this model comes from `intermediate' mass
satellites that orbit long enough in the cluster potential to be affected
significantly by stellar stripping.

Similar results have been found in other studies. In the work by
\citet{purcell07} mentioned above, the ICL on the cluster mass scale is largely
produced by the disruption of satellite galaxies with mass $\sim 10^{11}\,{\rm
  M}_{\odot}$. In a more recent work, \citet{martel12} combine N-body
simulations with a subgrid treatment of galaxy formation. 
They find that about 60 per cent of the ICL in
haloes more massive than $\sim 10^{14}\,{\rm M}_{\odot}$ is due to the
disruption of galaxies with stellar mass in the range $6\times 10^{8}
-3\times10^{10}\,{\rm M}_{\odot}$. 
Their results are close to predictions from our model Cont. Strip., with an
enhanced contribution of intermediate-mass galaxies with respect to the other
two models discussed in this study and results from \citet{purcell07}. 
However, in agreement with our results and those from \citet{purcell07}, 
they also find that the contribution from low-mass galaxies to the
ICL is negligible, though they dominate the cluster galaxy population in
number.

\begin{figure}
\begin{center}
\includegraphics[scale=.43]{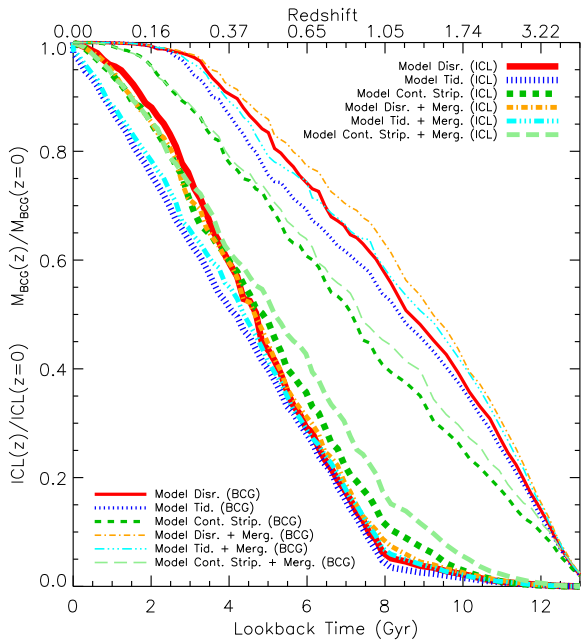}
\caption{Fraction of stellar mass in ICL as a function of the look-back time,
  normalized to the amount of ICL at present (thick lines of different style
  and color). Thinner lines show the evolution in the stellar mass of the main
  progenitor of BCGs as a function of lookback time.}
\label{fig:ICLwhen}
\end{center}
\end{figure}

The analysis discussed above answers the question on `which' galaxies
contribute (most) to the ICL component. We can now take advantage of results
from our models to ask `when' the ICL is produced. We address this issue in
Figure~\ref{fig:ICLwhen} that shows the ICL fraction (normalized to the total
amount of ICL measured at present) as a function of cosmic time, for all
different prescriptions used in this study (thick lines of different colour and
style). The cosmic evolution of the ICL component is compared with the
evolution of the stellar mass in the main progenitor of the corresponding BCGs,
shown as thin lines.

In agreement with previous studies both based on simulations
\citep{willman,giuseppe} and on analytic or semi-analytic models
\citep{conroy,pigi}, we find that the bulk of the ICL forms relatively late,
below $z=1$. Models Disr. and Tid. predict very similar ICL growth histories,
while in model Cont. Strip. the ICL formation appears to be anticipated with
respect to the other two models. At redshift $\sim 1$, less than 10 per cent of
the ICL was already formed in models Disr. and Tid. If the merger channel for
the formation of the ICL is switched on in these models, the ICL fraction
formed at the same redshift increases to $\sim 15-20$ per cent. As explained
earlier, the importance of the merger channel is enhanced in model Cont.
Strip. The figure also shows that the ICL grows slower than the mass in the main
progenitor of the BCG down to $z\sim 1$, i.e. at a lookback time of $\sim
8$~Gyr. Below this redshift, the ICL component grows much faster than the BCG,
with more than 80 per cent of the total ICL mass found at z=0 being formed
during this redshift interval.

\begin{figure}
\begin{center}
\includegraphics[scale=.60]{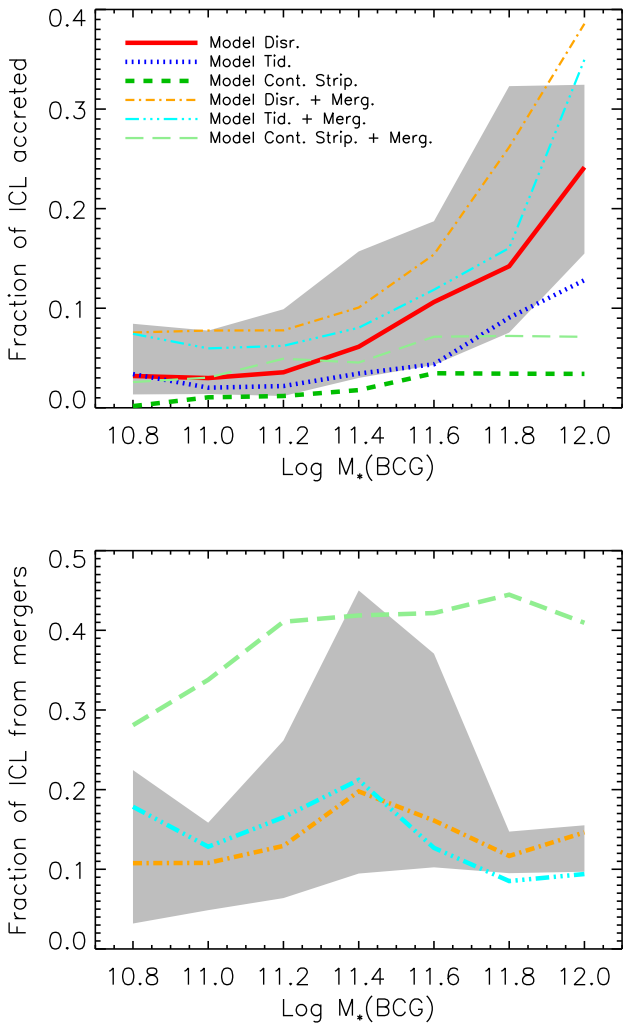}
\caption{Top panel: fraction of accreted ICL as a function of the BCG stellar
  mass.  Bottom panel: fraction of ICL contributed by the merger channel as a
  function of the BCG stellar mass. The grey shaded region shows the 20th and
  80th percentiles of the distribution measured for model Disr. The other
  models exhibit similar dispersions.}
\label{fig:ICLaccr}
\end{center}
\end{figure}

We now want to quantify what is the fraction of ICL that is accreted onto the
cluster during its hierarchical growth. As explained in
Section~\ref{sec:model}, the amount of ICL associated with a central galaxy
can increase through three channels: 
\begin{itemize}
  \item[(i)] stripping of satellite galaxies orbiting in the same parent halo;
  \item[(ii)] mergers, if this particular channel is switched on;
  \item[(iii)] accretion of the ICL component that is associated with new
    galaxies infalling onto the cluster during its assembly history or with
    satellite galaxies (i.e. when a Type 1 becomes a Type 2 galaxy in model
    Disr. or when a Type 1 is stripped for the first time in models Tid. and
    Cont. Strip).
\end{itemize}
In the following, we define the `accreted' component as the ICL fraction that
is coming through the third channel described above. The contribution from this
component is shown in the top panel of Figure~\ref{fig:ICLaccr} as a function
of the stellar mass of the BCG. In models Disr. and Tid., the fraction of the
accreted component increases from a few per cent for the least massive BCGs in
our sample to $\sim 13-25$ per cent for the most massive BCGs, in case the
merger channel is off. For model Cont. Strip., the increase as a function of the
BCG stellar mass is less pronounced, and the fraction of accreted ICL is always
below 10 per cent even in the case the merger channel is on.

The bottom panel of Figure~\ref{fig:ICLaccr} quantifies the amount of ICL that
comes from the merger channel. If this channel is switched on, as we have seen
in the left panel of Figure~\ref{fig:ICLfrac}, the ICL fraction increases in
each model.  We note that the amount of ICL that comes from this channel cannot
be inferred precisely by comparing each model in Figure~\ref{fig:ICLfrac} with
its counterpart including the merger channel, because it affects slightly the
merger times of galaxies. The contributions to the ICL coming from mergers are
shown in the bottom panel of Figure~\ref{fig:ICLaccr}, and have been stored
using the three prescriptions used in this study with the merger channel on. We
find that in models Disr. and Tid. the merger channel contributes to $\sim 15$
per cent of the total ICL. For model Cont. Strip., the contribution from mergers
is significantly larger, ranging from $\sim 30$ per cent for the least massive
BCGs in our sample, to $\sim 40$ per cent for the most massive ones.

\section[]{ICL and BCG properties}
\label{sec:bcgprop}

We now focus on the relation between the ICL and the main properties of BCGs,
such as stellar mass and luminosity, and analyse how these are affected by the
inclusion of our prescriptions for the formation of the ICL.

\begin{figure}
\begin{center}
\includegraphics[scale=.46]{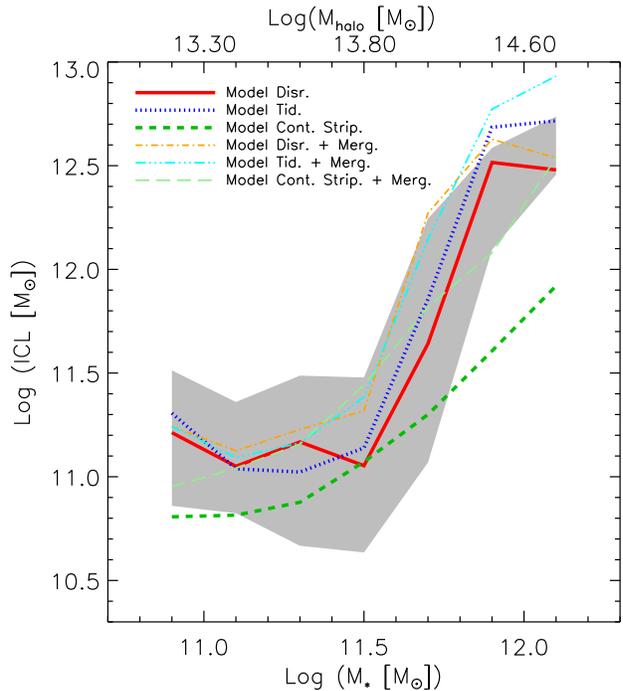}
\caption{Stellar mass in the ICL component as a function of the BCGs stellar
  mass. Lines of different style and colour correspond to different models, as
  indicated in the legend. The grey shaded region shows the 20th and 80th
  percentiles of the distribution obtained for model Disr.}
\label{fig:ICLBCGmass}
\end{center}
\end{figure}

In Figure \ref{fig:ICLBCGmass} we show the relation between the mass in the ICL
component and the stellar mass of the BCG. As expected, more massive BCGs
reside in haloes that host a more conspicuous ICL component. For models
Disr. and Tid., the correlation is strong for BCGs more massive than $\sim
3\times10^{11}\,{\rm M}_{\odot}$, while it is very weak for less massive
central galaxies. Model Cont. Strip. predicts a weaker correlation over the mass
range explored, and a significantly lower mass in the ICL component with
respect to the other two models when the merger channel is off.

\begin{figure}
\begin{center}
\includegraphics[scale=.45]{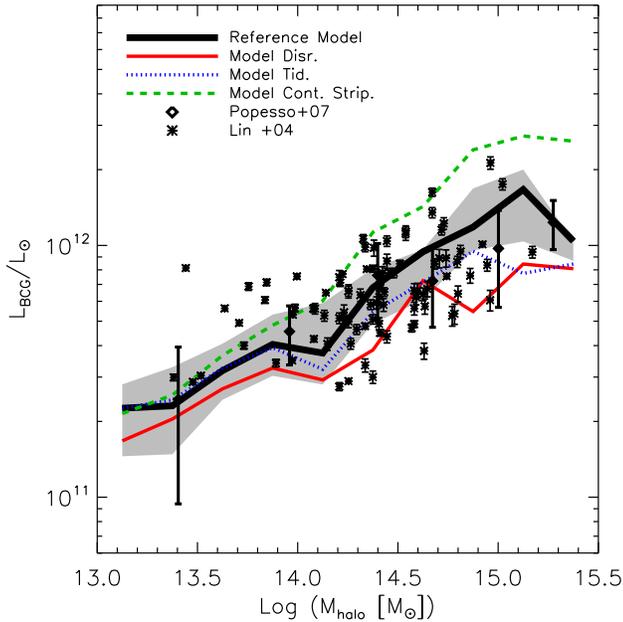}
\caption{Relation between the BCG luminosity in K-band and the cluster mass
  ($M_{200}$). Black stars with error bars show the observational
  measurements by \citet{lin}, while black diamonds with error bars are observational
  measurements by \citet{popesso07}.  The thick black solid line shows
  predictions from our reference model (DLB07), while thinner lines of
  different style show predictions from the models including our different
  prescriptions for the formation of the ICL. The grey shaded region shows the
  20th and 80th percentiles of the distribution obtained for the reference
  model. Our other models have comparable scatter.}
\label{fig:BCGlumhalomass}
\end{center}
\end{figure}

Figure~\ref{fig:BCGlumhalomass} shows the luminosity of the BCG in the K-band
as predicted by our models as a function of the halo mass. Model predictions
are compared with observational measurements by \citet{lin} and
\citet{popesso07}. It is worth recalling that our model luminosities are
`total' luminosities, that are difficult to measure observationally:
\citet{popesso07} use SDSS `model magnitudes', while \citet{lin} use elliptical
aperture magnitudes corresponding to a surface brightness of $\mu_K = 20 \,
\rm{mag/arcsec^2}$, and include an extra correction of 0.2 mag to get their
`total magnitudes'. Our reference model is in very good agreement with both
sets of observational data. Models Disr. and Tid. predict slightly lower
luminosities than our reference model, as a consequence of the reduced
accretion of stellar mass from satellites (either because part of these are
destroyed - model Disr., or because they are stripped - model Tid.). Model Cont.
Strip. predicts luminosities of central galaxies brighter than those measured by
Popesso et al., on the cluster mass scale. This is due to the fact that, as
mentioned earlier, merger times are shorter in this model, which increases the
stellar mass and the luminosity of the BCGs by accretion of (massive)
satellites.

\begin{figure}
\begin{center}
\includegraphics[scale=.85]{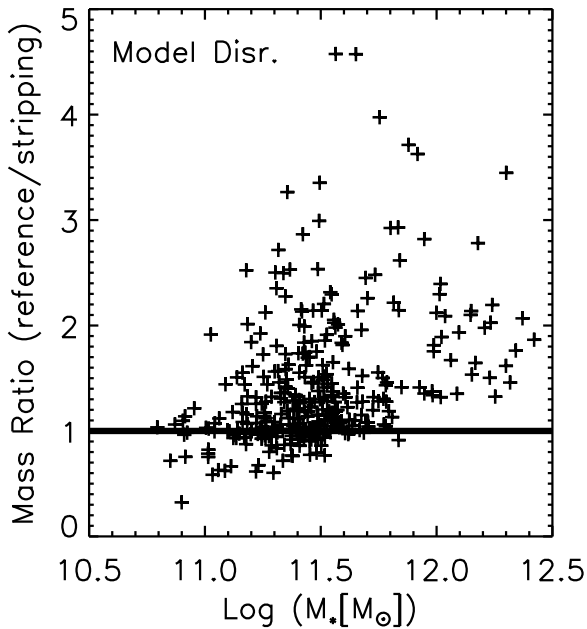}
\caption{Ratio between the stellar mass of BCGs in the reference model and the
  corresponding value in model Disr., as a function of the stellar mass in
  the reference model.}
\label{fig:bcgmass}
\end{center}
\end{figure}

Surprisingly, we find that also in models Disr. and Tid. a small fraction of
the BCGs are brighter (more massive) than in the reference model.  Naively, we
do not expect this to be possible as the only effect of our prescriptions
should be that of reducing the stellar mass of satellite galaxies by stripping
or disruption. In Figure~\ref{fig:bcgmass}, we show the ratio between the BCG
stellar mass in the reference model and the corresponding stellar mass in model
Disr., as a function of the former quantity. The Figure shows that the majority
of the BCGs are more massive in the reference model, but about 23 per cent of
the BCGs in our sample are actually {\it more} massive when we switch on our
prescription for the disruption of satellite galaxies. If we additionally
switch on the merger channel for the formation of the ICL, the fraction of BCGs
that are more massive in model Disr. than in the reference model reduces to
about 10 per cent. Model Tid. behaves in a similar way (the corresponding
fractions are 28 and 9 per cent, respectively). Model Cont. Strip., as also
evident from Figure~\ref{fig:BCGlumhalomass} behaves differently. In this
model, about half (42 per cent) of the BCGs are more massive than in the
reference model, with this fraction reducing to about 26 per cent when the
merger channel is switched on. While in model Disr. (and Tid.) the effect seems
to be limited to the less massive BCGs (those with stellar mass lower than
$\sim 10^{11.5} M_{\odot}$), in model Cont. Strip. this happens for BCGs of any
mass.

\begin{figure}
\begin{center}
\includegraphics[scale=.55]{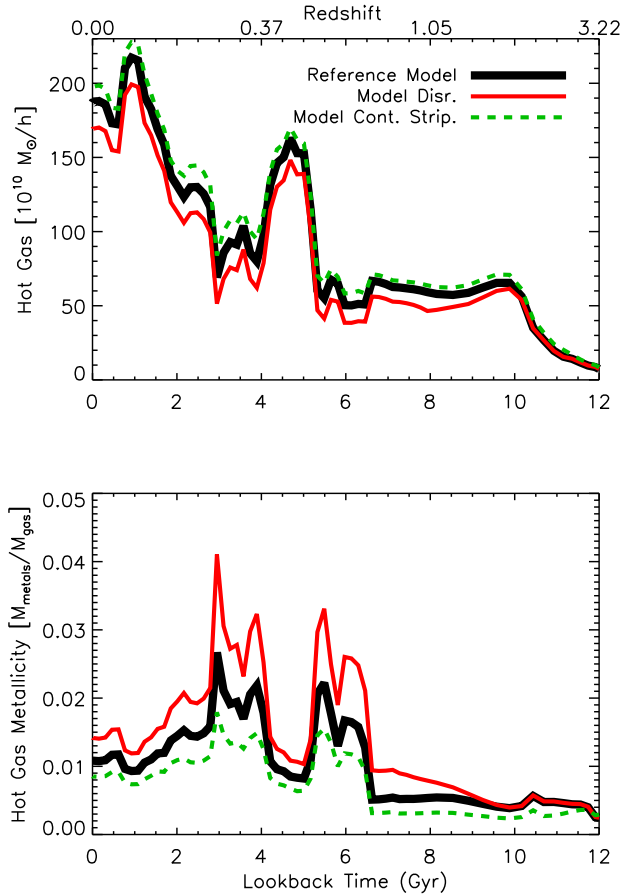}
\caption{Top panel: Mass in the hot component available at each redshift, for a
  BCG that is found to be more massive in model Disr. (and less massive in
  model Cont. Strip.)  than in the reference model. Bottom panel: hot gas
  metallicity at each redshift. Results are shown for different models, as
  indicated in the legend.}
\label{fig:BCGevol}
\end{center}
\end{figure}

In order to understand this finding, we have analysed the evolution of the
stellar mass, mass gained through mergers, and cooling rate for a number of the
BCGs that are more massive in model Disr. than in the reference model. We have
found that in all cases, the reason for the increased mass of the BCGs can be
traced back to a more efficient cooling rate. In order to illustrate this, we
show in Figure~\ref{fig:BCGevol} a representative example. The top panel shows
the amount of hot gas available for cooling onto the main progenitor of the
BCG, up to $z\sim 3$, while the bottom panel shows the corresponding
metallicity. Results are shown for the reference model, and for our models
Disr.  and Cont. Strip. (model Tid. behaves similarly to the reference
model). For this particular BCG, the evolution of both the hot gas content and
that of its metallicity in model Disr. follow very closely the evolution in the
reference model. At $z\sim 1.5$, the metallicity of the hot gas in model Disr.
becomes higher than in the reference model, and this causes a significant
increase in the cooling rate. In turn, the more efficient cooling determines an
increase in the star formation, and therefore of the final BCG stellar
mass. The behaviour is different in model Cont. Strip. where the metallicity of
the hot gas actually falls below the corresponding value in the reference
model. As we have explained earlier, for this particular model we have used a
different implementation of the merger times which introduces a net decrease in
the merger time of satellite galaxies. This more rapid merger rates causes
about half of the BCGs to be more massive in model Cont. Strip. than in the
reference model.

The difference in the hot gas metallicity of the BCG between model Disr. and
Cont. Strip. is due to a different metal content of the gas accreted from
satellite galaxies.  We recall that when a satellite galaxy is destroyed (or
stripped), its stellar mass goes to the ICL while its gaseous content,
including the corresponding metals, go to the hot gas component associated with
the central galaxy. In model Cont. Strip., stripping is a continuous process
that starts as soon as a galaxy becomes a satellite. In addition, in this model
the formation of the ICL component starts earlier than in the other models, as
shown in Figure~\ref{fig:ICLwhen}. In this case, the gas that is removed from
satellites tends to dilute the metallicity of the hot gas component. In model
Disr. (as well as in model Tid.), the formation of the ICL starts a bit later,
and satellite galaxies are more massive than in model Cont. Strip. (and
therefore also more metal rich - see Figure~\ref{fig:ICLstarmass}), so that
their disruption tends on average to increase the metal content of the hot gas
component.

\section[]{ICL metallicity}
\label{sec:metallicity}
 
From the observational viewpoint, little is known about the stellar populations
of the ICL component. Using I-band HST data, \citet{durrell02} compared
the brightness of Virgo ICL red giant branch (RGB) stars to that of RGB stars
in a metal-poor dwarf galaxies, and estimated an age for the ICL population
older than $\sim 2$~Gyr, and a relatively high metallicity ($-0.8\lesssim [{\rm
    Fe/H}] \lesssim -0.2$). \citet{williams07} use HST observations of a single
intra-cluster field in the Virgo Cluster and find that the field is dominated
by low-metallicity stars ($[{\rm M/H}]\lesssim -1.$) with ages older than $\sim
10$~Gyr. However, they find that the field contains stars of the full range of
metallicities probed ($-2.3 \leq [{\rm M/H}] \leq 0.0$), with the metal-poor
stars exhibiting more spatial structure than metal-rich stars, suggesting that
the intra-cluster population is not well mixed. Using long-slit spectra and
measuring the equivalent width of Lick indices, \citet{Coccato_etal_2011} find
that most of the stars in the dynamically hot halo of NGN3311 (the BCG in the
Hydra I cluster) are old and metal-poor ($[{\rm Z/H}]\sim -0.35$). 

In this section, we present predictions of our models concerning in particular
the metallicity of the ICL component. We recall that our model adopts an
instantaneous recycling approximation for chemical enrichment. In particular,
we assume that a constant yield of heavy elements is produced per solar mass of
gas converted into stars, and that all metals are instantaneously returned to
the cold phase. Metals are then exchanged between the different phases
proportionally to the mass flows. When a satellite galaxy is stripped of some
fraction of its stars (or destroyed), a proportional fraction (or all) of the
metals are also moved from the satellite stars into the ICL.

\begin{center}
\begin{figure*}
\includegraphics[scale=.48]{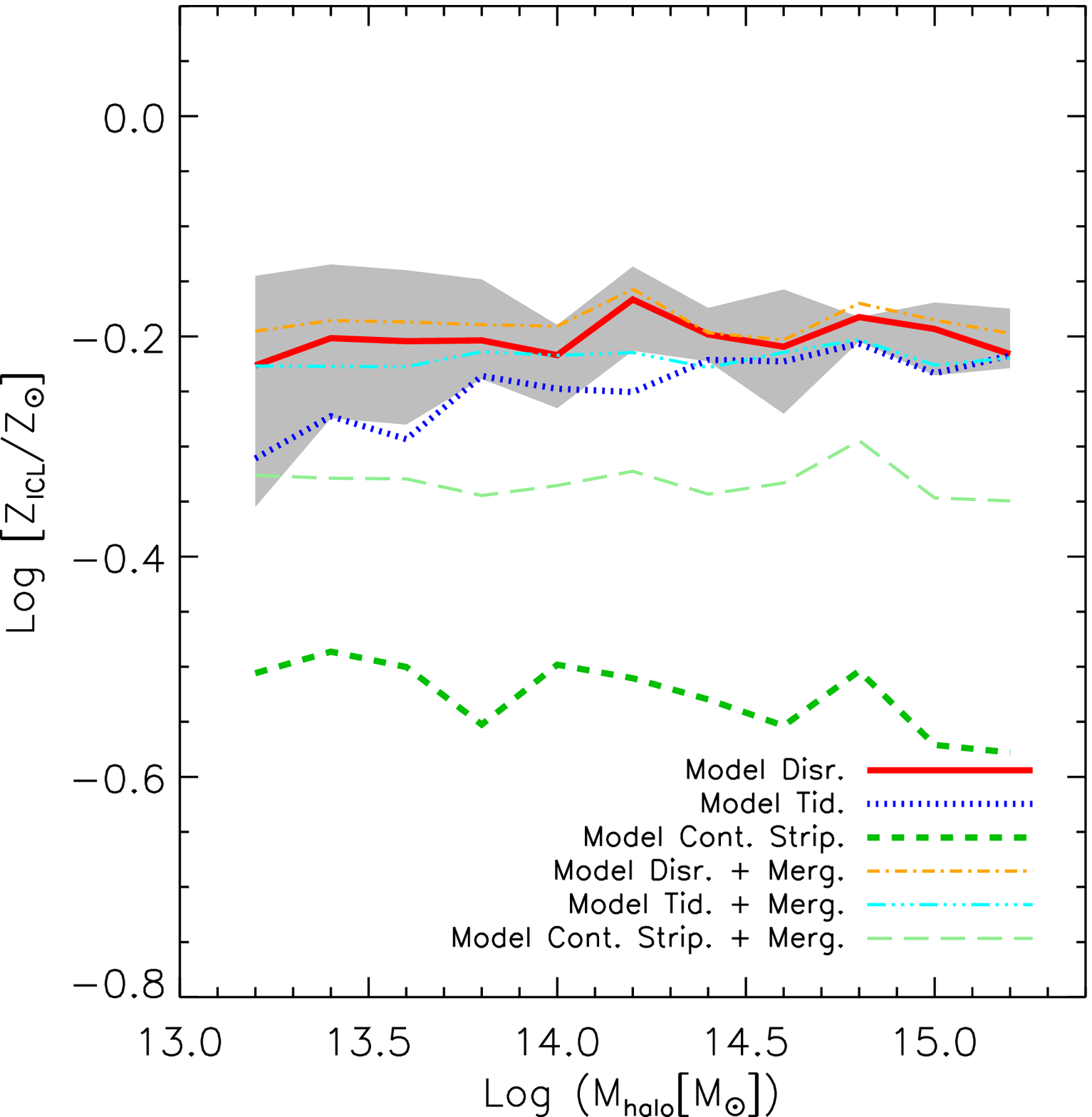}
\hspace{10pt}
\includegraphics[scale=.48]{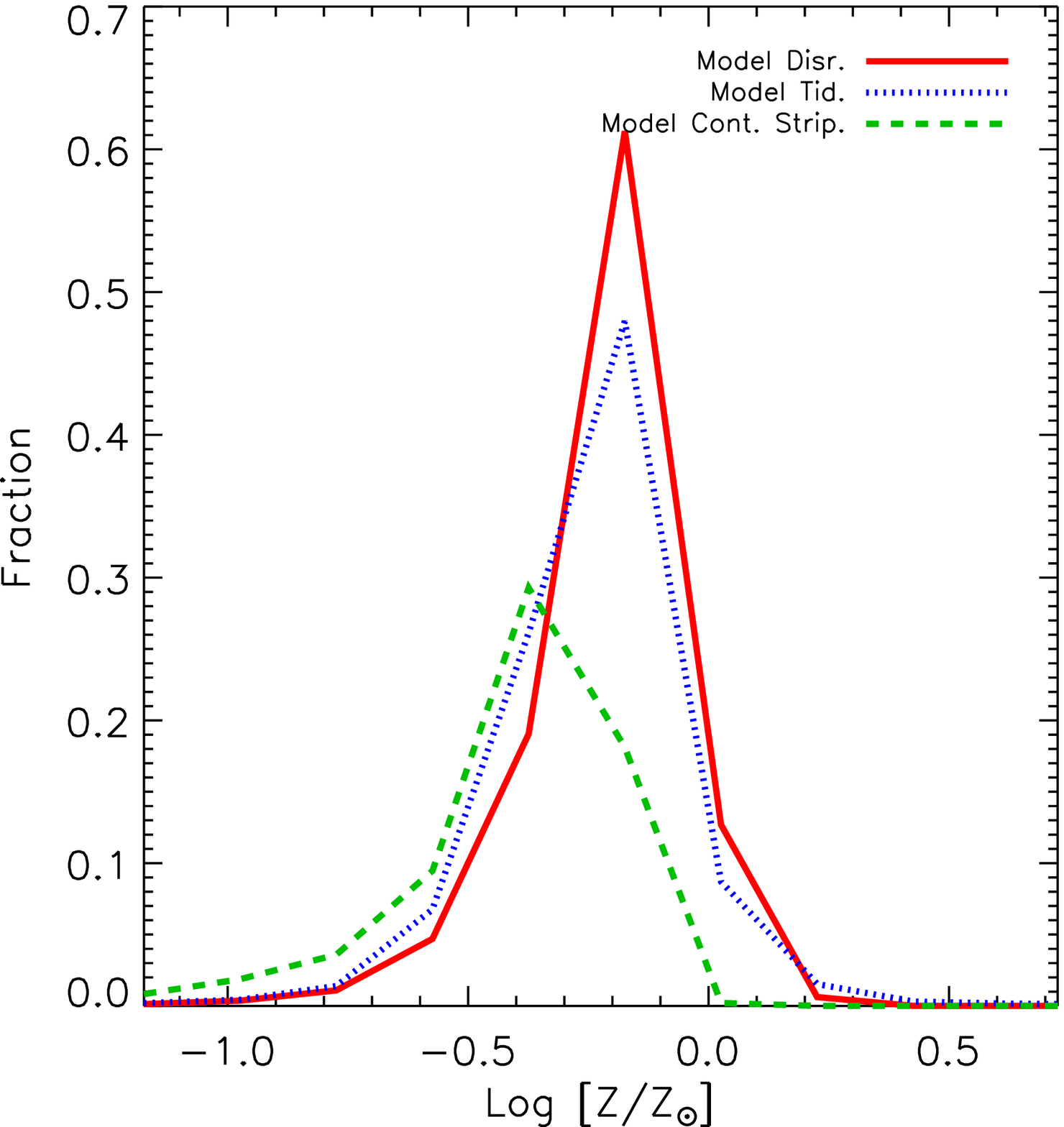}
\caption{Left panel: mean metallicity of the ICL component as a function of halo
  mass. Predictions from different models are shown using lines of different
  style and colour. The shaded gray region shows the 20th and 80th percentiles
  of the distribution obtained for model Disr. Right panel: average
  distribution of the metallicities of the stars in the ICL component for all
  haloes in our sample.}
\label{fig:ICLmet}
\end{figure*}
\end{center}

The left panel of Figure~\ref{fig:ICLmet} shows the median metallicity of the
ICL component, as a function of halo mass, for our different prescriptions. The
grey shaded region shows the 20th and 80th percentiles of the distribution
found for model Disr. (the other models exhibit similar dispersions).  In our
models, the ICL metallicity does not vary significantly as a function of the
halo mass. Assuming $Z_{\odot}=0.02$, the average metallicity of the ICL 
in models Disr. and Tid. is $ \sim 0.63 Z_{\odot}$, while for model Cont. Strip.
the ICL metallicity is significantly lower ($\sim 0.32-0.50 Z_{\odot}$).  This
is a consequence of the fact that the galaxies contributing to the ICL are on
average less massive (and more metal-poor) than those that contribute to the
ICL in models Disr. and Tid. (see Figure~\ref{fig:ICLstarmass}). In the right
panel of Figure~\ref{fig:ICLmet}, we show the average metallicity distribution of stars
in the ICL component for all haloes in our sample. As a consequence of the
results shown in Figure~\ref{fig:ICLstarmass}, model Cont. Strip. predicts a
distribution shifted towards lower metallicities, with a peak at $\sim 0.4
Z_{\odot}$. Models Disr. and Tid. predict distributions that are less broad and
peaked at higher metallicities.

Our model results are therefore qualitatively consistent with observational measurements by 
\citet{williams07}, with most of the stars in the ICL having sub-solar
metallicity but covering a relatively wide range.

\begin{center}
\begin{figure}
\includegraphics[scale=.45]{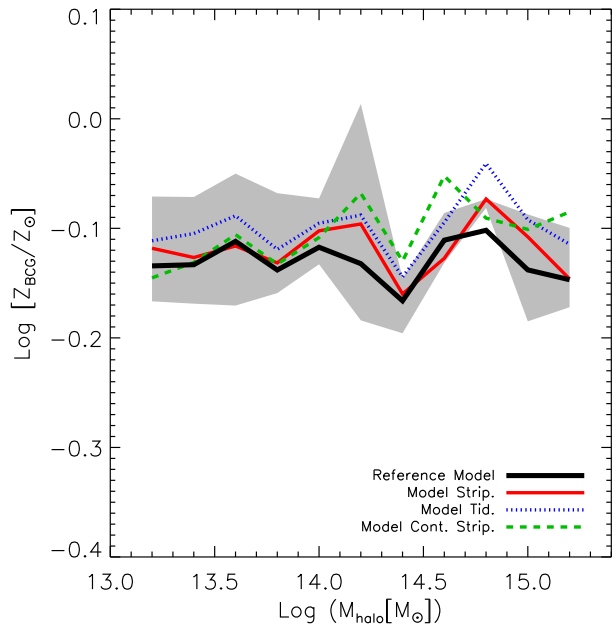} 
\caption{BCG stellar metallicity as a function of halo mass. Lines of different
  style and colour correspond to the different prescriptions used in our study,
  as indicated in the legend. The thick solid black line shows predictions from
  our reference model.}
\label{fig:BCGmet}
\end{figure}
\end{center}

Figure~\ref{fig:BCGmet} shows the BCG metallicity predicted by the different
models used in our study, as a function of the halo mass. The figure shows that
all models predict very similar metallicities for the BCGs, of about $ 0.74 \,
Z_{\odot}$.  Predictions are close to those of the reference model (thick solid
black line). On average, stellar stripping slightly increases the BCG
metallicity, particularly in model Tid. This happens because the most massive
galaxies receive less low-metallicity stars from satellite galaxies whose
masses (and metal contents) are reduced because of stellar stripping.

Our results confirm findings by \citet{gabste} who show that the model
mass-metallicity relation is offset low with respect to the observational
measurements at the massive end. In particular, the observed BCG metallicities
(similar to those of the most massive galaxies) are expected to be at least
0.2-0.3~dex larger \citep{vonderlinden07,loubser09}. Figure~\ref{fig:BCGmet}
shows that the inclusion of a model for the formation of the ICL component does
not significantly improve this disagreement. More in general, we find that our
modelling of the formation of the ICL component does not
significantly affect the predicted mass-metallicity
relation. This is in apparent contrast with findings by \citet{henriques} who
claim that the introduction of satellite disruption is sufficient to bring the
stellar metallicities of the most massive galaxies in agreement with the
observational data. We note that \citet{henriques} use the same reference model
adopted in our study and employ a Monte Carlo Markov Chain parameter estimation
technique to constrain the model with the K-band luminosity function, the B-V
colours, and the black hole-bulge mass relation. The `best fit' model found by
\citet{henriques} includes a model for tidal stripping of the satellite
galaxies, but also adopts different parameters with respect to the reference
model, in particular for the supernovae feedback and gas recycling process (see
their Table 2).  We therefore argue that, as discussed also in \citet{gabste},
that stellar stripping cannot provide alone the solution to the problem
highlighted above, and that modifications of the star formation and feedback
processes are required.

\section[]{Discussion and conclusions}
\label{sec:discussion}

In this work, we build upon the semi-analytic model presented in \citet[][
  DLB07]{dlb} to describe the generation of intra-cluster light (ICL). We 
include different implementations for modelling the formation
of the diffuse ICL. In particular, we consider: (i) a
model that assumes the stellar component of satellite galaxies can be affected
only after their parent dark matter substructures are stripped below the
resolution limit of the simulation (Disruption model); (ii) a model that
accounts for stellar stripping also from satellites sitting in distinct dark
matter subhaloes, and based on a simple estimate of the tidal radius (Tidal
Radius model); and (iii) a model based on a fitting formula derived from a
suite of numerical simulations aimed to study the evolution of a disk galaxy
within the global tidal field of a group environment \citep[][ Continuous
  Stripping model]{alvaro}. In addition, we have also considered the relaxation
processes acting during galaxy-galaxy mergers by simply assuming that 20 per
cent of the stellar mass of the merging satellite gets unbound and ends-up in
the ICL component associated with the remnant galaxy. In our implementations,
the bulk of the ICL is produced through tidal stripping and disruption of the
satellite galaxies, with the merger channel contributing only for a minor
fraction. 

The reference model we have used is known to over-predict the abundance of
galaxies with mass below $\sim 10^{10}\,{\rm M}_{\odot}$
\citep{Fontanot_etal_2009,qiguo}. The inclusion of a model for stellar
stripping of satellite galaxies alleviates this problem, but does not solve it.
In a recent work, \citet{budzynski12} use a catalogue of groups and clusters 
from SDSS DR7 in the redshift range $0.15 \leq z \leq 0.4$ and compare the galaxy number 
density profiles with predictions from the same reference model used in our study. 
They show that the model follows very well the observational measurements but in 
the very central regions (within $\sim 0.2 R_{500}$), where the predicted profile is steeper 
than observational measurements. The inclusion of stellar stripping would improve the 
agreement with data in this region where the tidal field is stronger and galaxies 
are more likely to be stripped. However, the same comparison with the data by 
\citet{lin2} would lead to an opposite conclusion.  This suggests that the uncertainty on the 
density profiles in the inner region is probably too large to put strong constraints 
on stripping models.

As we discuss below, we find that the dominant contribution to the ICL
formation comes from stripping and/or disruption of massive satellites so that
the excess of intermediate to low-mass galaxies does not affect significantly
our results. In our Cont. Strip. model, we use a different prescription for
merger times with respect to that employed in the reference model and in the
other two models considered (see Section \ref{sec:modc}). As a consequence, merger times
are on average shorter than in the other models which result in an
under-prediction of massive satellites. We note that recent work by
\citet{Villalobos_etal_2013} has pointed out that the dynamical friction formula 
used in our Continuous Stripping model (\citet{bk}) under-estimate merger times 
at higher redshift. Implementing their proposed modification would make merger 
times longer in this model, making results more similar to the other two models. 
We have explicitly tested (by adding a
fudge factor that increases again the merger times so as to bring the
predicted mass function in agreement with observations) that this does not
affect significantly the results presented in this work.

A number of recent studies have focused on the formation of the ICL, using both
hydrodynamical numerical simulations
\citep[e.g.][]{giuseppe,puchwein,rudick06}, and analytic models based on
subhalo infall and evolution \citep[e.g.][]{purcell07,watson}. Less work on the
subject has been carried out using semi-analytic models of galaxy formation,
but for basic predictions in terms of how the fraction of ICL depends on the
parent halo mass \citep{pigi,somerville,qiguo}.

Our models predict an ICL fraction that varies between $\sim 20$ and $\sim 40$
per cent (depending on the particular implementation adopted), with no
significant trend as a function of the parent halo mass. Results are in
qualitative agreement with observational data, in particular on the cluster
mass scale. 
We note, however, that the ICL fractions predicted by our models depend on 
the resolution of the simulations: for a set of simulations that use a particle 
mass one order of magnitude larger than that adopted in the high resolution runs 
used in our study, the predicted ICL fractions increase by ~30-40 per cent.
We stress that both the data and the model predictions exhibit a
relatively large halo-to-halo scatter. On the cluster mass scale, we find that
the scatter is largely due to a variety of mass accretion histories at fixed
halo mass, as argued by \citet{purcell07}: objects that formed earlier (that
were also more concentrated) had more time to strip stars from their satellite
galaxies or accumulate ICL through accretion of smaller systems. On group
scale, where the predicted scatter is very large, we do not find any clear
correlation between ICL fraction and halo concentration or formation time. We
show that, on these scales, large part of the scatter is driven by individual
accretion events of massive satellites. The (albeit weak) correlation between
the ICL fraction and concentration on the cluster mass scale can be tested
observationally, e.g. by measuring the ICL fraction and concentration for 
system lying in a relatively narrow halo mass range. 

Our models predict that the ICL forms very late, below redshift $z\sim 1$, in
agreement with previous analysis based on hydrodynamical numerical simulations
\citep[e.g.][]{giuseppe}. About 5 to 25 per cent of the diffuse light has been
accreted during the hierarchical growth of dark matter haloes, i.e. it is 
associated with new galaxies falling onto the haloes during their
assembly history.  In addition, we find that the bulk of the ICL is produced by
the most massive satellite galaxies, $M \sim 10^{10-11}\,{\rm M}_{\odot}$, in
agreement with recent findings based on N-body simulations \citep{martel12} and
analytic models \citep{purcell07}. Low-mass galaxies ($M_* < 10^{9} M_{\odot}$)
contribute very little to the ICL in terms of mass, although they dominate
in terms of number. This is a natural consequence of dynamical friction 
triggering the generation of the ICL: the
most massive satellites approach the inner cluster regions faster than their
less massive counterparts. Close to the cluster centre, tidal forces are
stronger, increasing the stripping efficiency. In contrast, small satellites
spend most of their time in the outer regions where tidal stripping is weaker.

Since most of the ICL is produced by tidal stripping of massive satellites,
this component is found to have a metallicity that is similar to that of these
galaxies. Our model predictions are in qualitative agreement with observations,
with most of the stars in the diffuse component having on average sub-solar
metallicities (but covering a relatively large range). We also find that the
mean metallicity of the ICL is approximately constant as a function of halo
mass, and exhibit a relatively small halo-to-halo scatter. In contrast,
\citet{purcell08} predict a weak increase of the ICL metallicity with increasing 
halo mass, over the same halo mass range considered in our study. 
Finally, we show that the inclusion of a model for tidal stripping of satellite 
galaxies does not significantly affect the predicted mass-metallicity relation,
and only slightly increases the metallicity of the most massive galaxies. For all
models, these galaxies have stellar metallicities significantly lower than observed 
(see also \citealt{gabste}).
Future and more detailed observations focused e.g. on age and metallicity of the
ICL component will help constraining our models and understanding the physical 
mechanisms driving the formation of this important stellar component.

\section*{Acknowledgements}
EC, GDL and AV acknowledge financial support from the European Research Council
under the European Community's Seventh Framework Programme (FP7/2007-2013)/ERC
grant agreement n. 202781. This work has been supported by the PRIN-INAF 2009
Grant ``Towards an Italian Network for Computational Cosmology'', the PRIN-MIUR
2009 grant ''Tracing the Growth of Structures in the Universe'' and the PD51
INFN grant. Simulations have been carried out at the CINECA National
Supercomputing Centre, with CPU time allocated through an ISCRA project and an
agreement between CINECA and University of Trieste.  We acknowledge partial
support by the European Commissions FP7 Marie Curie Initial Training Network
CosmoComp (PITN-GA-2009-238356). We thank the referee, Stefano Zibetti, Magda 
Arnaboldi and Douglas Watson for useful comments that helped us improving our 
manuscript.

\bsp

\label{lastpage}


\bibliographystyle{mn2e}
\bibliography{biblio}

\appendix

\section{Numerical Convergence}
\label{sec:numconv}

\begin{center}
\begin{figure*}
\includegraphics[scale=.45]{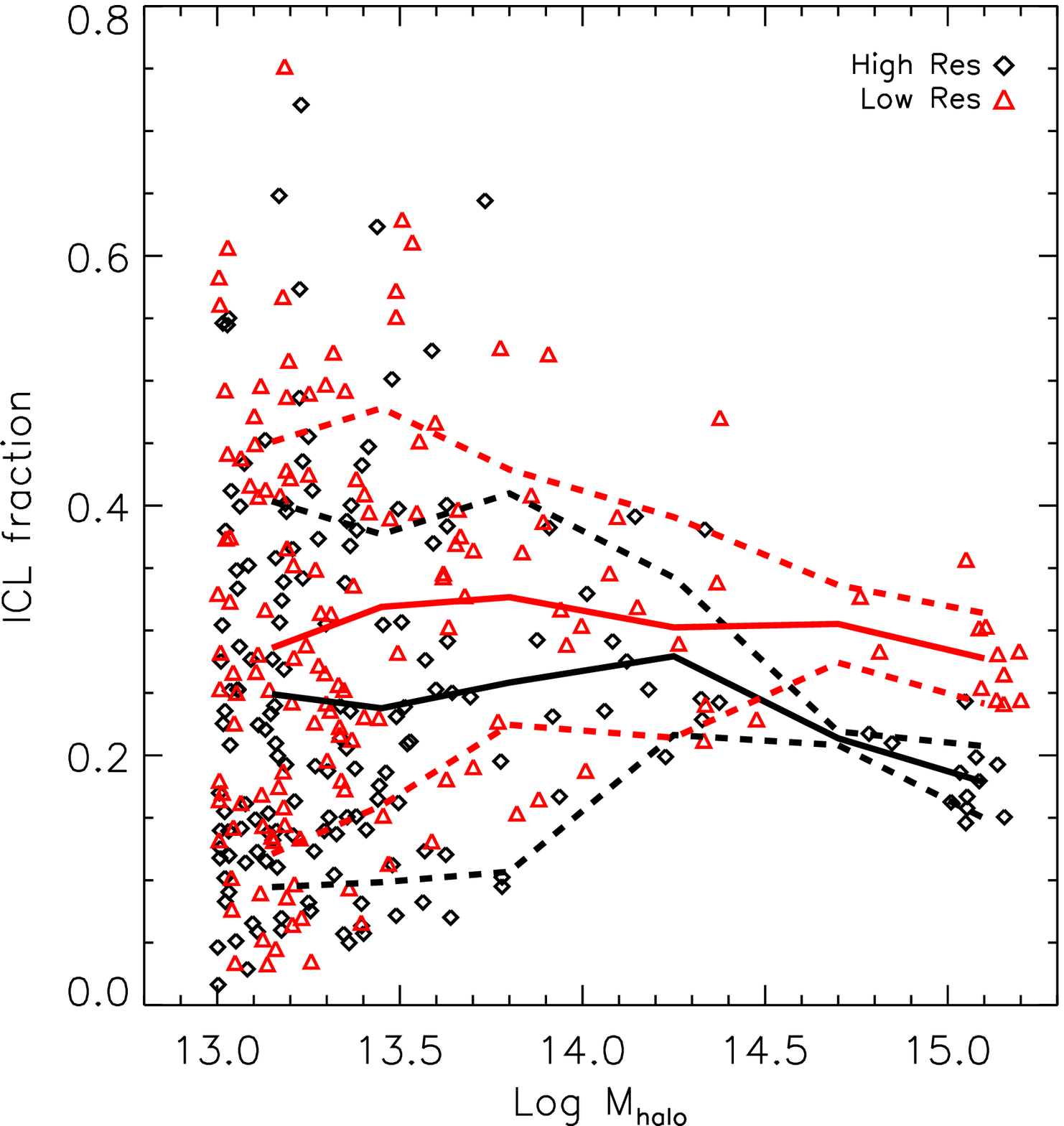}
\hspace{20pt}
\includegraphics[scale=.45]{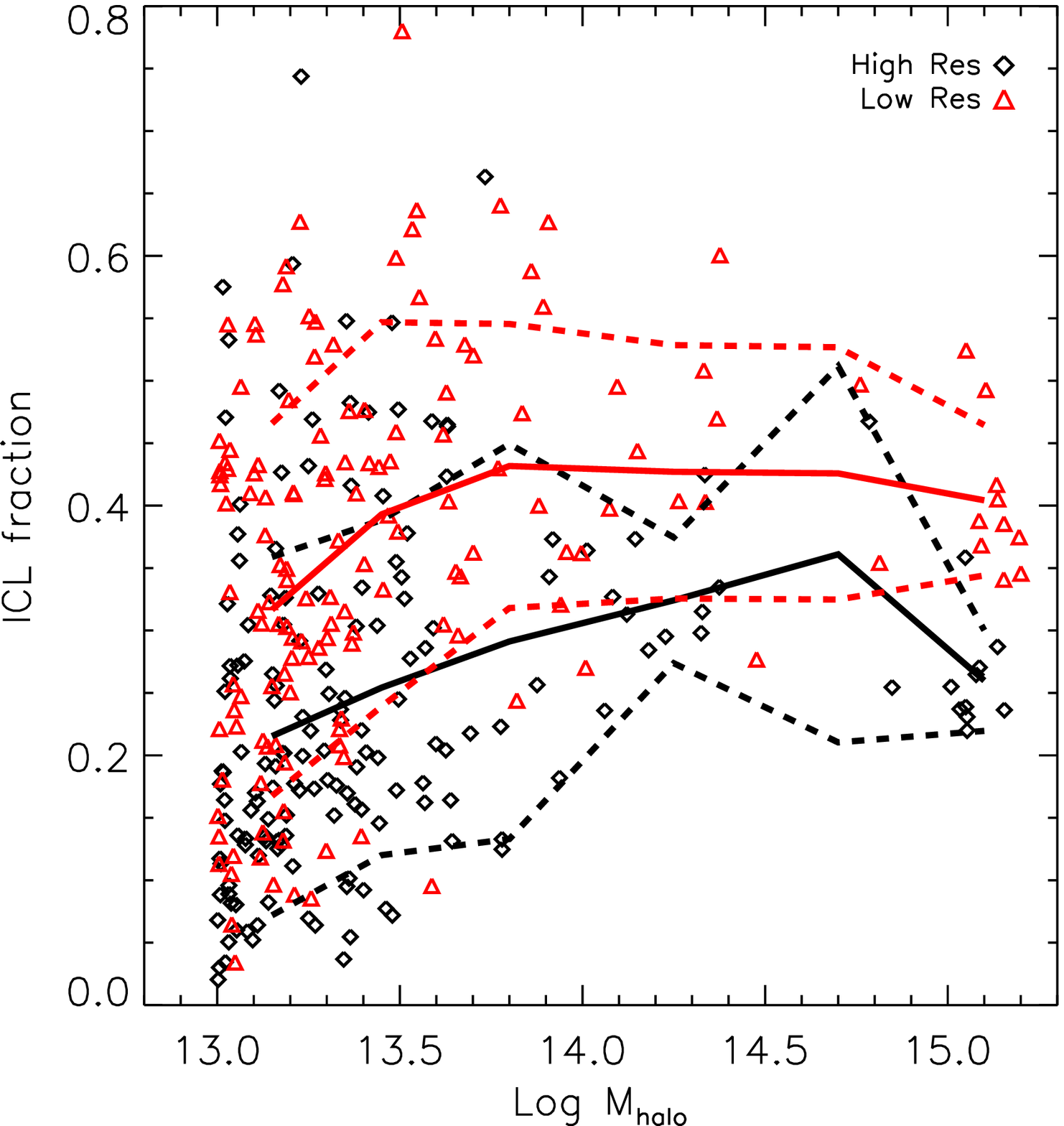}
\caption{Left Panel: ICL fraction predicted by the Disruption model as a function of halo
mass, for haloes in the high resolution set used in this paper (black lines and
symbols), and the corresponding low-resolution version (red lines and symbols).
Dashed lines show the one sigma dispersion, while solid lines correspond to the
mean. Right Panel: same as the left panel, but for the Tidal Radius model.}
\label{fig:ICLconv}
\end{figure*}
\end{center}
In the left panel of Figure \ref{fig:ICLconv} we show the ICL fraction as a 
function of cluster mass, as predicted by the Disruption model applied to two sets 
of simulations: a set with high-resolution (the one we used in the paper), and the 
corresponding low-resolution set where the same initial conditions were used for each halo
but the particle mass adopted is one order of magnitude larger. The ICL fraction
appears to be systematically higher in the low-resolution set over all the
halo mass range considered, with the difference being more significant on cluster
scale. This is due to the fact that, decreasing the resolution, a larger fraction of
satellite galaxies are classified as Type 2 and are subject to our stripping model.
The effect is weaker in low-mass haloes that have a lower number of satellites
galaxies. 

We carried out the same convergence test also for Tidal Radius model, and show 
the corresponding results in the right panel of Figure \ref{fig:ICLconv}. 
The ICL fraction is higher in the low-resolution set than in the high resolution set, 
by about 40 per cent. Due to the constraint given by equation \ref{eqn:eq_radii}, 
events of stripping of Type 1 satellites are extremely rare in this model, and 
their contribution to the ICL fraction is lower than 1 per cent over all the halo mass 
range considered. This means that Type 2 galaxies are the main contributors to the 
total amount of ICL.
The larger fraction of these galaxies in the low-resolution set causes the increase
of the ICL fraction, also on group scale, where the increase seems to be more
important than it is in Disruption model.

\end{document}